\documentclass[12pt]{article}
\pdfoutput=1
\usepackage[nosort]{cite}

\usepackage{epsfig}
\usepackage{amsfonts}
\usepackage{amscd}
\usepackage{latexsym}
\usepackage{amsmath,amssymb,mathtools}
\usepackage{verbatim}
\usepackage{setspace}
\usepackage[dvipsnames]{xcolor}
\usepackage{fancyhdr}
\usepackage{cite}
\usepackage{hyperref}
\usepackage{xcolor}
\usepackage{tikz}
\usepackage{tikz-cd}
\usepackage{slashed}
\usepackage{multirow}
\usetikzlibrary{calc}
\usetikzlibrary{topaths}
\usetikzlibrary{decorations}
\usetikzlibrary{decorations.pathmorphing}
\usetikzlibrary{arrows,decorations.markings,cd}
\usetikzlibrary{calc,arrows,cd,decorations.markings,snakes}
\tikzset{
->-/.style args={#1rotate#2}{decoration={markings, mark=at position #1 with {\arrow[scale=1.5,rotate = #2 ]{stealth}}}, postaction={decorate}}
}
\usetikzlibrary{shapes.geometric}
\usetikzlibrary{knots}
\usepackage{tikz-3dplot}
\usepackage{tabularray}

\tikzset{snake it/.style={decorate, decoration=snake}}

\usepackage{draft}
\usepackage{hyperref}
\usepackage{graphicx,color,subcaption}
\usepackage{cite}
\usepackage{mciteplus}
\usepackage{skak}
\usepackage{bbm}
\usepackage[english]{babel}
\usepackage{amsthm}
\usepackage{soul}
\usepackage{makecell}

\numberwithin{equation}{section}

\def\bZ{\mathbb{Z}}

\usepackage{bbding}

\def\({\left(}
\def\){\right)}

\def\f{\phi}
\def\al{\alpha}
\def\om{\omega}
\def\lam{\lambda}
\def\we{\wedge}
\newcommand{\ishi}[1]{|#1\rangle\!\rangle}

\usepackage{pifont}

\begin{document}

\begin{titlepage}

\title{On the Absence of Symmetric Simple Conformal Boundary Conditions}

\author{Pengcheng Wei$^1$ and Yunqin Zheng$^2$}

        \address{${}^{1}$ Department of Physics, Nanjing University, Nanjing, Jiangsu 210093, China\\\bigskip 
        ${}^{2}$ Kavli Institute for Theoretical Sciences, \\University of Chinese Academy of Sciences, Beijing, 100190, China
        }

\abstract

Non-trivial 't Hooft anomaly obstructs the existence of a simple symmetric conformal boundary condition in a CFT. Conversely, there is a common piece of lore that trivial 't Hooft anomaly promises the existence of a simple symmetry conformal boundary condition in a given CFT. Recently, counter examples to this lore was realized in tetracritical Ising CFT \cite{Choi:2023xjw} and compact boson \cite{SMGtalk}---the simple conformal boundary conditions preserving certain anomaly-free subsymmetry are absent in these CFTs.  In this work, we uncover the underlying reason for the absence of these boundary conditions in counter examples, and propose a criterion diagnosing when the lore fails for any given 2d CFT. The Symmetry TFT description for boundary conditions plays a crucial role.

\end{titlepage}

\eject

\setcounter{tocdepth}{3} 

\tableofcontents

\bigskip

\section{Introduction}\label{sec: intro}

Anomalous symmetries provide powerful constraints on the dynamics of quantum field theories and quantum many-body systems. In particular, any theory with an anomalous symmetry must have nontrivial features at low energies---the system can not have a non-degenerate, symmetric, ground state in the thermodynamical limit. Recently, this property has been adopted as a definition of anomaly \cite{Thorngren:2019iar,Chang:2018iay,Choi:2023xjw}. 

On the contrary, starting with a CFT defined on the spacetime $\Sigma$ with some \emph{anomaly-free} symmetry, it has been conjectured that there exists a symmetric deformation which drives the CFT to a symmetry preserving trivially gapped phase\footnote{Trivially gapped means being gapped with one ground state on arbitrary spatial manifold.}, dubbed symmetric mass generation \cite{SeibergString2019Talk, Mouland:2025ilu, Wang:2022ucy, SMGtalk, Razamat:2020kyf, Tong:2021phe, Tong:2019bbk, You:2017ltx, Lu:2022qtc}.\footnote{A relevant deformation may not always exist in a given CFT. More generally, the deformation can consist of ``flowing up" to a mother theory, and flowing down in a different direction while preserving the symmetry. An example of flowing up and down is discussed in \cite{Gaiotto:2019asa}. } The trivially gapped phase means 
the energy spectrum at the IR fixed point has a unique vacuum/ground state with a finite gap in the thermodynamic limit. 
Such gapping deformation has been constructed explicitly for some theories in literature \cite{Haldane:1988zza, Wang:2013yta, Han:2017hdv, Zeng:2022grc, Wang:2022ucy, Wang:2018ugf, vanBeest:2023dbu}.

The 't Hooft anomaly also constraints the existence of simple symmetric conformal boundary conditions of the CFTs. In particular, it has been shown that non-trivial 't Hooft anomaly obstructs the existence of such conditions \cite{Chen:2023hmm, Han:2017hdv, Li:2022drc}.

Alternatively, when a CFT has an anomaly-free symmetry, one can turn on a symmetric gapping deformation on half space. Flowing the entire system to the deep infrared, the gapping deformation generates a symmetric conformal boundary condition for the CFT. Although the boundary condition is not guaranteed to be simple by this construction, it has been tested that simple symmetry conformal boundary conditions do exist in many 2d examples, including free fermions with multiple copies of $U(1)$ symmetries \cite{Han:2017hdv, Li:2022drc, Smith:2019jnh, Smith:2020nuf}, and Wess-Zumino-Witten models with various gauge groups and levels with invertible center symmetries \cite{Li:2022drc, Numasawa:2017crf}. There are also examples in higher dimensions  \cite{Thorngren:2020yht}. Moreover, examples including Ising$^2$ and permutation orbifold CFTs with fusion category symmetries are also discussed recently in \cite{Choi:2023xjw}. This motivates a common piece of lore in boundary CFT \cite{BoyleSmith:2022duw}\footnote{In this work, we will only consider CFTs in two spacetime dimensions. }
\paragraph{Lore:}  For any compact 2d CFT $Q$ with an internal fusion category symmetry $\mathcal{C}$, there exists a simple conformal boundary condition that preserves any anomaly-free subsymmetry $\mathcal{D}$. 
\\ 

According to \cite{Choi:2023xjw}, when the fusion category $\mathcal{D}$ is non-invertible, $\mathcal{D}$ can be either \emph{strongly} anomaly free or \emph{weakly} anomaly free. The former is equivalent to being gaugable with a maximal Frobenius algebra $A= \oplus_{a\in \mathcal{D}} d_a a$ where $d_a$ is the quantum dimension of $a$,\footnote{It turns out that strongly anomaly free is also equivalent to being compatible with a trivially gapped phase, which is more commonly stated in the literature.  } while the latter is equivalent to being gaugable with a possibly non-maximal Frobenius algebra $A= \oplus_{a\in \mathcal{D}} n_a a$ with $n_a$ a positive integer.   Correspondingly, the simple conformal boundary condition can be \emph{strongly} or \emph{weakly} symmetric---the former implies the existence of symmetric boundary state in the closed string channel, while the latter implies the existence of symmetric ground state in the open string channel. Incorporating this subtlety, there are two versions of the lore, namely, the strong and weak versions.

However, several counterexamples to the lore are also known in the literature. For example, it is pointed out in \cite{Choi:2023xjw} that, in the unitary minimal model $M(6,5)$ with diagonal modular invariance, there does not exist any conformal boundary state strongly preserving the anomaly-free $\operatorname{Rep}(S_3)$ symmetry, as a subcategory of its Verline lines. Yet another puzzle \cite{SMGtalk} is that people have not known any compact conformal boundary conditions of the compact boson at generic irrational radius $R$ preserving the $\mathbb{Z}_p\times\mathbb{Z}_q\subset U(1)^m\times U(1)^w$ symmetry, which can be shown to be anomaly-free.\footnote{A non-compact boundary condition preserving the symmetry has been found in \cite{Janik:2001hb}. Since it is non-compact, it is more subtle (if not impossible) to require simpleness. Hence we focus on compact boundary conditions throughout this paper. }

The main goal of this paper is to explain why the lore fails for these CFTs, and provide general criteria for when the lore can fail in a given CFT. It turns out that describing the boundary conditions in the Symmetry TFT (SymTFT) framework \cite{Choi:2024tri, Choi:2024wfm, Bhardwaj:2024igy, Cordova:2024iti, Cordova:2024vsq, Cordova:2024goh,Copetti:2024onh,Copetti:2024dcz} is powerful in answering these questions. With the criteria clarified, we propose an improved piece of lore in Section \ref{sec: criteria}. 

We also point it out the converse statement of the lore is valid, which has already been proposed in literature such as \cite{Han:2017hdv, Li:2022drc,Thorngren:2020yht}.

The paper is organized as follows. In Section \ref{sec: symtft}, we review some SymTFT results, especially focusing on the boundary state, which will serve as the main technique we use throughout the paper. Section \ref{sec: criteria} is our main result, in which we propose the criteria for the existence of a symmetric simple conformal boundary condition. Then we apply our criteria to several examples in detail, which are the compact boson in Section \ref{sec: boson}, several minimal models in Section \ref{sec: min model}
and the $SU(3)_3$ and the $SU(2)_1\times SU(2)_3\times SU(2)_{-4}$ WZW models in Section \ref{sec: su3 wzw}.

\subsection*{Notation Guide}

We summarize here the categorical notations used throughout this work.

\begin{center}
\begin{tabular}{rp{0.78\textwidth}}
$\mathcal{C}, \mathcal{D}, \cdots$ &
A fusion category of symmetry lines operators/defeats. \\

$\mathcal{Z}(\mathcal{C})$ &
The Drinfeld center of a fusion category $\mathcal{C}$. \\

$\mathcal{B}$ &
A topological boundary condition of the SymTFT $\mathcal{Z}(\mathcal{C})$; or the fusion category of line operators (anyons) supported on this boundary. (See footnotes \ref{footnote1} and \ref{footnote2}.) \\

$F_{\mathcal{B}}$ &
The bulk-to-boundary forgetful functor $F_{\mathcal{B}}: \mathcal{Z}(\mathcal{C}) \to \mathcal{B}$. \\

$\mathcal{A}$ &
A Lagrangian algebra in the bulk SymTFT $\mathcal{Z}(\mathcal{C})$. \\

$\mu \prec \mathcal{A}$ &
The object $\mu\in\mathcal{Z}(\mathcal{C})$ in the Lagrangian algebra object $\mathcal{A}$. \\

$A$ &
A symmetric, separable, special, haploid Frobenius algebra in the symmetry fusion category $\mathcal{C}$. \\

$\mathcal{C}_A$ &
The category of right $A$-modules in $\mathcal{C}$. \\
\end{tabular}
\end{center}

\section{Symmetry TFT for Boundary States}\label{sec: symtft}

In this section, we review previous results on the SymTFT construction of the boundary state of a 2d CFT. See \cite{Choi:2024tri, Choi:2024wfm, Bhardwaj:2024igy, Cordova:2024iti, Cordova:2024vsq, Cordova:2024goh,Copetti:2024onh,Copetti:2024dcz} for more details. This framework will serve as the primary tool for analyzing the symmetry properties of simple conformal boundaries.

\subsection{SymTFT for a CFT}\label{sec: symtftcft}
SymTFT is a framework in which symmetry-related properties and manipulations of a field theory can be treated systematically. 

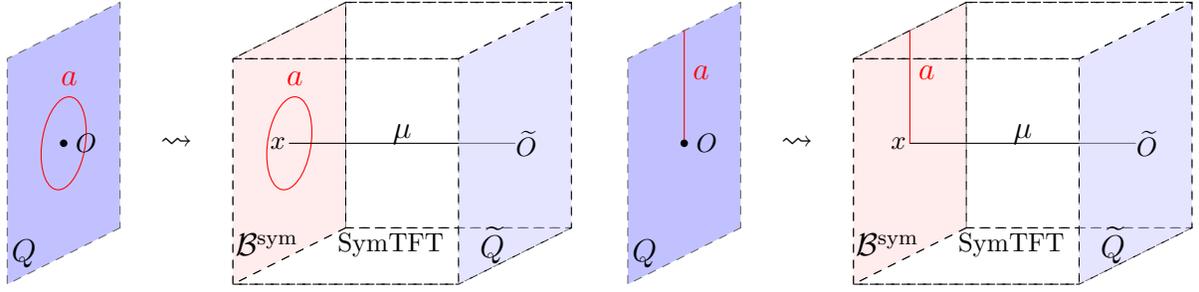
\begin{figure}
\centering
\begin{tikzpicture}[
    x={(-0.5cm,-0.25cm)},  
    y={(1cm,0cm)},         
    z={(0cm,1cm)},         
    scale=1.5,
    line join=round,
    line cap=round
]
\begin{scope}
\draw[fill=red!15,opacity=0.5,dashed] (0,0,0) -- (2,0,0) -- (2,0,2) -- (0,0,2) -- cycle;
\begin{scope}[canvas is xz plane at y=0]
    \draw[red] (1,1) circle (0.4cm);
\end{scope}
\draw (1,0,1) -- (1,2,1);
\draw[dashed] (0,0,0) -- (2,0,0) -- (2,2,0) -- (0,2,0) -- cycle;
\draw[dashed] (0,0,0) -- (0,2,0) -- (0,2,2) -- (0,0,2) -- cycle;
\draw[dashed] (2,0,0) -- (2,0,2) -- (2,2,2) -- (2,2,0) -- cycle;
\draw[fill=blue!20,opacity=0.5,dashed] (0,2,0) -- (2,2,0) -- (2,2,2) -- (0,2,2) -- cycle;
\draw[dashed] (0,0,2) -- (2,0,2) -- (2,2,2) -- (0,2,2) -- cycle;
\draw[fill=blue!40,opacity=0.6,dashed] (0,-2,0) -- (2,-2,0) -- (2,-2,2) -- (0,-2,2) -- cycle;
\begin{scope}[canvas is xz plane at y=-2]
    \draw[red] (1,1) circle (0.4cm);
\end{scope}
\node at(1,-1,1){$\rightsquigarrow$};
\node at(1.7,-2,0.2){$Q$};
\node at(1.4,0,0.2){$\mathcal{B}^{\text{sym}}$};
\node at(1.4,2,0.2){$\widetilde{Q}$};
\node at(1.4,1.1,0.2){\footnotesize{SymTFT}};
\node[circle, fill=black, inner sep=1pt] at (1,-2,1) {};
\node at(1,-0.1,1){\footnotesize{$x$}};
\node at(1,2.1,1){\footnotesize{$\widetilde{O}$}};
\node at(1,1,1.08){$\mu$};
\node[red] at(0.7,-0.1,1.5){$a$};
\node[red] at(0.7,-2.1,1.5){$a$};
\node at(1,-1.8,1){\footnotesize{$O$}};
\end{scope}

\begin{scope}[xshift=5.5cm]
\draw[fill=red!15,opacity=0.5,dashed] (0,0,0) -- (2,0,0) -- (2,0,2) -- (0,0,2) -- cycle;
\draw (1,0,1) -- (1,2,1);
\draw[dashed] (0,0,0) -- (2,0,0) -- (2,2,0) -- (0,2,0) -- cycle;
\draw[dashed] (0,0,0) -- (0,2,0) -- (0,2,2) -- (0,0,2) -- cycle;
\draw[dashed] (2,0,0) -- (2,0,2) -- (2,2,2) -- (2,2,0) -- cycle;
\draw[fill=blue!20,opacity=0.5,dashed] (0,2,0) -- (2,2,0) -- (2,2,2) -- (0,2,2) -- cycle;
\draw[dashed] (0,0,2) -- (2,0,2) -- (2,2,2) -- (0,2,2) -- cycle;
\draw[fill=blue!40,opacity=0.6,dashed] (0,-2,0) -- (2,-2,0) -- (2,-2,2) -- (0,-2,2) -- cycle;
\draw[red] (1,0,1) -- (1,0,2);
\draw[red] (1,-2,1) -- (1,-2,2);
\node at(1,-1,1){$\rightsquigarrow$};
\node at(1.7,-2,0.2){$Q$};
\node at(1.4,0,0.2){$\mathcal{B}^{\text{sym}}$};
\node at(1.4,2,0.2){$\widetilde{Q}$};
\node at(1.4,1.1,0.2){\footnotesize{SymTFT}};
\node[circle, fill=black, inner sep=1pt] at (1,-2,1) {};
\node at(1,-0.1,1){\footnotesize{$x$}};
\node at(1,2.1,1){\footnotesize{$\widetilde{O}$}};
\node at(1,1,1.08){$\mu$};
\node at(1,-1.8,1){\footnotesize{$O$}};
\node[red] at(0.5,-0.1,1.5){$a$};
\node[red] at(0.5,-2.1,1.5){$a$};
\end{scope}

\end{tikzpicture}
\caption{Symmetry TFT for 2d CFT without boundaries.}
\label{symtft cft}
\end{figure}

Let us start with a 2d CFT $Q$ with generalized symmetry described by the fusion category $\mathcal{C}$ defined on a closed 2-manifold $\Sigma$. The theory $Q$ can be decomposed into a triplet (the left of Figure \ref{symtft cft})
\begin{align}
    Q \rightsquigarrow  (\mathcal{B}^{\text{sym}}, \text{SymTFT}, \widetilde{Q}).
\end{align}
SymTFT is the 3d Turaev-Viro TQFT, denoted as $\mathrm{TV}_\cC$, of the input fusion category $\cC$ defined on $\Sigma\times [0,1]$, whose line operators are described by the Drinfeld center of $\mathcal{C}$, i.e. $\mathcal{Z}(\mathcal{C})$. The symmetry boundary $\mathcal{B}^{\text{sym}}$ is a particular topological boundary condition of SymTFT, on which supporting line operators are labeled by $\mathcal{C}$, i.e. $\mathcal{B}^{\text{sym}}\simeq\mathcal{C}$.\footnote{\label{footnote1} In our work, we do not distinguish between a (topological) boundary condition of 3d SymTFT, and the category of (topological) lines (defects/operators) supported on it, which we both denote as $\mathcal{B}$. Note that the notation here is different from \cite{Choi:2024tri}, where $\mathcal{B}\simeq \mathcal{C}_A$ is a $\mathcal{C}$-module category. But here $\mathcal{B}\simeq {}_A\mathcal{C}_A$ denotes the category of the anyons support on the boundary, which are the dual symmetry lines after gauging $A$.} For any topological boundary condition $\mathcal{B}$ of SymTFT, we use $F_{\mathcal{B}}:\mathcal{Z}(\mathcal{C})\to\mathcal{B}$ to denote the forgetful functor from the bulk $\mathcal{Z}(\mathcal{C})$ to boundary $\mathcal{B}$.  The physical boundary $\widetilde{Q}$ is a possibly non-topological boundary condition of SymTFT, depending on the dynamics detail of $Q$. 

A point operator $O$ in $Q$ is expanded into (see the left of Figure \ref{symtft cft})
\begin{align}
    O \rightsquigarrow (x, \mu, \widetilde{O}).
\end{align}
Here $\mu$ is a line in the SymTFT, which could condense on $\mathcal{B}^{\text{sym}}$; $x$ is an operator in the topological junction space of $\mu$ on $\mathcal{B}^{\text{sym}}$, which we denote as $x \in W_\mu=\operatorname{Hom}_{\mathcal{B}^{\text{sym}}}(F_{\mathcal{B}^{\text{sym}}}(\mu), 1)$; and $\widetilde{O}$ is an operator in the junction space of $\mu$ on the physical boundary $\widetilde{Q}$, denoted as $\widetilde{O}\in \mathcal{V}_\mu$.

Since symmetry lines $a, b,c\cdots$ in $\mathcal{C}$ are supported on the symmetry boundary $\mathcal{B}^{\text{sym}}$, the symmetry actions---visualized as shrinking the circle surrounding the junction $x$ (the red line in Figure \ref{symtft cft} left) ---act as linear maps from $W_\mu$ to itself. Thus, the bulk line $\mu$ serves as the label of the "representation" carried by $O$. 

More generally, an operator $O$ in $a$-twisted Hilbert space of CFT can be expanded into $(x, \mu, \widetilde{O})$ (Figure \ref{symtft cft} right), where $x$ is an operator in the space $W_\mu^a\coloneqq\operatorname{Hom}_{\mathcal{B}^{\text{sym}}}(F_{\mathcal{B}^{\text{sym}}}(\mu), a)$. The $a$-twisted Hilbert space of CFT admits such decomposition 
\begin{align}\label{hilb space decompose}
    \mathcal{H}^a=\bigoplus_{\mu}\mathcal{H}^a_\mu=\bigoplus_{\mu}W_\mu^a\otimes\mathcal{V}_\mu,
\end{align}
where the sum is over simple lines in $\mathcal{Z}(\mathcal{C})$.

\subsection{SymTFT for a Boundary CFT}\label{sec:SymTFTBCFT}

\begin{figure}
\centering
\begin{tikzpicture}[
    x={(-0.5cm,-0.25cm)},  
    y={(1cm,0cm)},         
    z={(0cm,1cm)},         
    scale=1.5,
    line join=round,
    line cap=round
]
\draw[fill=red!15,opacity=0.5,dashed] (0,0,0) -- (2,0,0) -- (2,0,2) -- (0,0,2) -- cycle;
\draw[dashed] (0,0,0) -- (2,0,0) -- (2,2,0) -- (0,2,0) -- cycle;
\draw[dashed] (0,0,0) -- (0,2,0) -- (0,2,2) -- (0,0,2) -- cycle;
\draw[dashed] (2,0,0) -- (2,0,2) -- (2,2,2) -- (2,2,0) -- cycle;
\draw[fill=blue!20,opacity=0.5,dashed] (0,2,0) -- (2,2,0) -- (2,2,2) -- (0,2,2) -- cycle;
\draw[fill=gray!40,opacity=0.8,dashed] (0,0,2) -- (2,0,2) -- (2,2,2) -- (0,2,2) -- cycle;
\draw[fill=blue!40,opacity=0.6,dashed] (0,-2,0) -- (2,-2,0) -- (2,-2,2) -- (0,-2,2) -- cycle;
\node at(1,-1,1){$\rightsquigarrow$};
\node at(1.7,-2,0.2){$Q$};
\node at(1.4,0,0.2){$\mathcal{B}^{\text{sym}}$};
\node at(1.4,2,0.2){$\widetilde{Q}$};
\node at(1.4,1.1,0.2){\footnotesize{SymTFT}};
\draw[very thick] (0,-2,2) -- (2,-2,2);
\draw[purple,opacity=0.8,very thick] (0,0,2) -- (2,0,2);
\draw[very thick] (0,2,2) -- (2,2,2);
\node at(1,1,1.99){$\mathcal{B}^{\text{bdy}}$};
\node at(1,-0.5,1.99){$\underline{B}$};
\node at(1,-2.5,1.99){$B$};
\node at(1,2.35,1.99){$\widetilde{B}$};
\draw[red] (0,0,1.2) -- (2,0,1.2);
\draw[red] (0,-2,1.2) -- (2,-2,1.2);
\node[red] at(0.7,-2.1,1){$a$};
\node[red] at(0.7,0.1,1){$a$};
\end{tikzpicture}
\caption{SymTFT for 2d boundary CFT.}
\label{symtft bdy cft}
\end{figure}
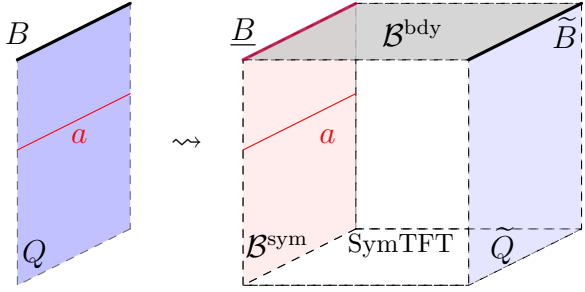

Now, consider the theory $Q$ defined on a 2-manifold with boundary, say the half-plane. The boundary condition $B$ can also be decomposed into a triplet (see Figure \ref{symtft bdy cft})
\begin{align}\label{bdy triplet}
    B\rightsquigarrow (\underline{B}, \mathcal{B}^{\text{bdy}}, \widetilde{B}).
\end{align}
Here, the 3d SymTFT is defined on a 3-manifold with corners, where we refer to $\underline{B}$ as the \emph{symmetry corner} and to $\widetilde{B}$ as the \emph{physical corner}.\footnote{\label{footnote2}For convenience, we do not distinguish between the geometric notions (such as boundaries and corners) and the corresponding relative physical theories (boundary conditions and corner conditions) placed on them. In this work, no confusion arises from this identification.} $\underline{B}$ is simple if it can not be decomposed into a direct sum of two symmetry corners between $\mathcal{B}^{\text{sym}}$ and $\mathcal{B}^{\text{bdy}}$.

The surface $\mathcal{B}^{\text{bdy}}$ is a topological boundary condition of the 3d SymTFT. The symmetry lines act on the symmetry corner $\underline{B}$ via parallel fusion (red line $a$ in Figure \ref{symtft bdy cft}), 
\begin{align}
    a\otimes \underline{B} = \bigoplus_{\underline{B}'} n_{a \underline{B}}^{\underline{B}'} \underline{B}',
\end{align}
where $n_{a \underline{B}}^{\underline{B}'}$ is a  non-negative integer matrix element. Thus the symmetry corner forms a non-negative integer matrix representation (NIM-rep). The topological boundary $\mathcal{B}^{\text{bdy}}$ serves as the label of the multiplet to which $B$ belongs, with respect to the symmetry $\mathcal{C}$, hence $\mathcal{B}^{\text{bdy}}$ serves as the label of the NIM-rep carried by the boundary condition $B$. 

The symmetry corner $\underline{B}$ is purely topological and is labeled by lines in the $(\mathcal{B}^{\text{sym}}, \mathcal{B}^{\text{bdy}})$-bimodule category. When $\mathcal{B}^{\text{sym}}$ and $\mathcal{B}^{\text{bdy}}$ are related by gauging an algebra object $A$ in $\mathcal{C}\simeq \mathcal{B}$, this description can be simplified to the category of right $A$-modules in $\mathcal{C}$.

The physical corner $\widetilde{B}$ is an interface between the boundary $\mathcal{B}^{\text{bdy}}$ and the physical boundary $\widetilde{Q}$. Throughout this work, we focus on the conformal boundaries of CFT's, hence the conformal condition should be imposed on the physical corner $\widetilde{B}$. We discuss this in the context of conformal boundary states in the following.

\subsection{SymTFT for a Boundary State}

The conformal boundary state is a state "in" the closed-string channel Hilbert space. Due to the decomposition \eqref{hilb space decompose}, we expect that a boundary state also admits such a decomposition. We shift the geometry from the half-plane to the annulus and focus on one boundary circle. 

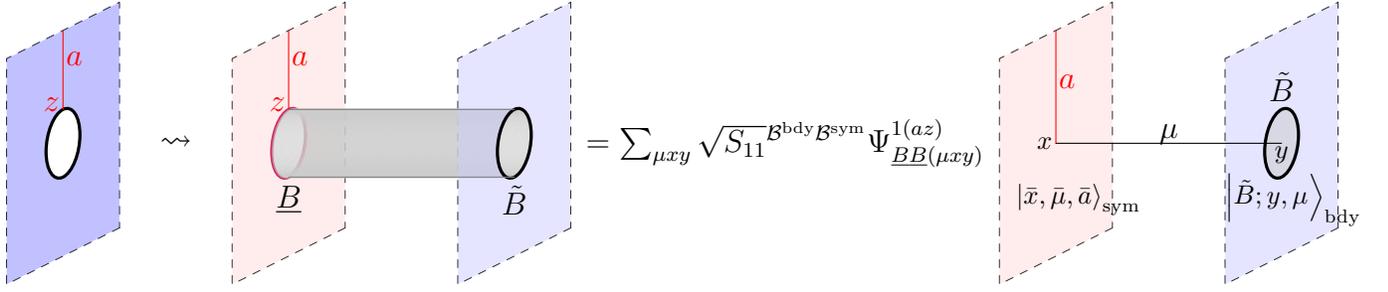
\begin{figure}
\centering
\begin{tikzpicture}[
    x={(-0.5cm,-0.25cm)},  
    y={(1cm,0cm)},         
    z={(0cm,1cm)},         
    scale=1.5,
    line join=round,
    line cap=round
]

\begin{scope}[xshift=-1cm]
    
\draw[dashed] (0,-2,0) -- (2,-2,0) -- (2,-2,2) -- (0,-2,2) -- cycle;
\begin{scope}[canvas is xz plane at y=-2]
    \fill[blue!40,opacity=0.6, even odd rule]
        (0,0) rectangle (2,2)
        (1,1) circle (0.3cm);
    \draw[very thick] (1,1) circle (0.3cm);
\end{scope}
\draw[red] (1,-2,1.3) -- (1,-2,2);
\node[red] at (0.8,-2,1.7) {$a$};
\node[red] at (1.2,-2,1.4) {$z$};

\node at (1,-1,1) {$\rightsquigarrow$};

\end{scope}

\begin{scope}[xshift=-1cm]

\draw[dashed] (0,0,0) -- (2,0,0) -- (2,0,2) -- (0,0,2) -- cycle;
\draw[dashed] (0,2,0) -- (2,2,0) -- (2,2,2) -- (0,2,2) -- cycle;

\begin{scope}[canvas is xz plane at y=0]
    \fill[fill=red!15,opacity=0.5, even odd rule]
        (0,0) rectangle (2,2)
        (1,1) circle (0.3cm);
    \draw[purple,opacity=0.8,very thick] (1,1) circle (0.3cm);
\end{scope}

\foreach \yy in {0,0.01,...,2} {
    \begin{scope}[canvas is xz plane at y=\yy]
        \fill[fill=gray!40,opacity=0.5, even odd rule]
            (1,1) circle (0.3cm)
            (1,1) circle (0.27cm);
    \end{scope}
}

\begin{scope}[canvas is xz plane at y=2]
    \fill[fill=blue!20,opacity=0.5, even odd rule]
        (0,0) rectangle (2,2)
        (1,1) circle (0.3cm);
    \draw[very thick] (1,1) circle (0.3cm);
\end{scope}

\draw[red] (1,0,1.3) -- (1,0,2);
\node[red] at (0.8,0,1.7) {$a$};
\node[red] at (1.2,0,1.4) {$z$};
\draw[gray] (1,0,1.3) -- (1,2,1.3);
\draw[gray] (1,0,0.7) -- (1,2,0.7);
\node at (1,0,0.5) {$\underline{B}$};
\node at (1,2,0.5) {$\tilde{B}$};

\end{scope}

\begin{scope}[xshift=1.8cm]

\node at (1,1.6,1) {$=\sum_{\mu x y} \sqrt{S_{11}}{}^{\mathcal{B}^{\text{bdy}}\mathcal{B}^{\text{sym}}}\Psi^{1(az)}_{\underline{B}\underline{B}(\mu xy)}$};

\draw[dashed] (0,4,0) -- (2,4,0) -- (2,4,2) -- (0,4,2) -- cycle;
\draw[dashed] (0,6,0) -- (2,6,0) -- (2,6,2) -- (0,6,2) -- cycle;

\begin{scope}[canvas is xz plane at y=4]
    \fill[fill=red!15,opacity=0.5]
        (0,0) rectangle (2,2);
\end{scope}

\begin{scope}[canvas is xz plane at y=6]
    \fill[fill=blue!20,opacity=0.5]
        (0,0) rectangle (2,2);
    \fill[fill=gray!40,opacity=0.5]
        (1,1) circle (0.3cm);
    \draw[very thick] (1,1) circle (0.3cm);
\end{scope}

\draw[red] (1,4,1) -- (1,4,2);
\node[red] at (0.8,4,1.5) {$a$};
\draw (1,4,1) -- (1,6,1);

\node at (1,3.9,1) {\footnotesize{$x$}};
\node at (1,6,0.9) {\footnotesize{$y$}};
\node at (1,5,1.08) {$\mu$};
\node at (1,4.2,0.5) {\footnotesize{$\ket{\bar{x},\bar{\mu}, \bar{a}}_{\text{sym}}$}};
\node at (1,6.1,0.5) {\footnotesize{$\ket{\tilde{B};y,\mu}_{\text{bdy}}$}};
\node at (1,6,1.5) {$\tilde{B}$};

\end{scope}

\end{tikzpicture}
    \caption{SymTFT for a boundary state}
    \label{fig:symtft bdy state}
\end{figure}

For future use, we need to consider an $a$-twisted boundary state as shown in Figure \ref{fig:symtft bdy state}, where a topological line $a$ terminates on the boundary $B$ at a topological junction $z$. It is well known that a boundary state $\ket{B}_{az}$ is a superposition of Ishibashi states with the coefficients constrained by modular invariant condition, a.k.a. Cardy condition. We will see the boundary state is realized in the Sandwich construction.

First, we perform the SymTFT blow-up, which is almost the same as before. We note that in the middle there is a hollow tube. Since the tube is topological, we can shrink it and move the boundary condition $\mathcal{B}^{\text{bdy}}$ to the right boundary. This results in a sum over anyon $\mu$ in the Lagrangian algebra $\mathcal{A}^{\text{bdy}}$ associated with $\mathcal{B}^{\text{bdy}}$ that terminates on $\mathcal{B}^{\text{sym}}$ at the topological junction $x$ and on $\mathcal{B}^{\text{bdy}}$ at the junction $y$, weighted by a coefficient determined by the half-braiding matrices $\Psi$. Further details can be found in \cite{Choi:2024tri}. 
Up to this point, every step is purely topological, involving only the data of the 3d SymTFT and its topological boundaries. In summary, the boundary state $\ket{B}$ can be expressed as a superposition of states analogous to those in \eqref{hilb space decompose}, which we analyze in detail below.

The state is a tensor product of two parts. On the left (symmetry boundary $\mathcal{B}^{\text{sym}}$) is a state in the topological junction space\footnote{Here, bar means $\mu$ is past oriented.}: 
\begin{align}
    \ket{\bar{x},\bar{\mu},\bar{a}}_{\text{sym}}\in W^a_\mu.
\end{align}
According to the decomposition \eqref{hilb space decompose}, the state on the right is supposed to be a state in $\mathcal{V}_\mu$. First, $\mu$ topologically terminates at the boundary $\mathcal{B}^{\text{bdy}}$, providing a state in $\operatorname{Hom}_{\mathcal{B}^{\text{bdy}}}\left(F_{\mathcal{B}^{\text{bdy}}}(\mu), 1\right)$. Then after the radial time evolution, this state is acted by the physical corner $\widetilde{B}$. Thus we define the physical corner $\widetilde{B}$ as the linear map:
\begin{align}
    \widetilde{B}: \operatorname{Hom}_{\mathcal{B}^{\text{bdy}}}\left(F_{\mathcal{B}^{\text{bdy}}}(\mu), 1\right) \to \mathcal{V}_\mu,
\end{align}
such that it's image satisfies certain conformal conditions and will be termed as \emph{half-Ishibashi state}, which will be explained in detail in Section \ref{sec: criteria}. 
We denote the state as\footnote{We point it out for completeness that, as the interface between two boundary conditions (i.e. $\mathcal{B}^{\text{bdy}}$ and $\widetilde{Q}$) of the bulk TQFT, $\widetilde{B}$ should also give raise to a well-defined $D^2$ Hilbert space of the TQFT. The space is denoted as $\mathcal{V}_{\alpha}^{\widetilde{B}\widetilde{B}}$ in \cite{Choi:2024tri}, for $\al\in\mathcal{B}^{\text{bdy}}$.
}
\begin{align}
    \ket{\widetilde{B};y,\mu}_{\text{bdy}}\coloneqq\widetilde{B}\left(\ket{y,\mu}_{\text{bdy}}\right)\in\mathcal{V}_\mu.
\end{align}

Putting ingredients together, we arrive at the formula for boundary state
\begin{align}\label{bdy state}
    \ket{B}_{a,z}=\sum_{\mu xy} \sqrt{\frac{S_{11}}{S_{1\mu}}} {}^{\mathcal{B}^{\text{bdy}}\mathcal{B}^{\text{sym}}}\Psi^{1(az)}_{\underline{B}\underline{B}(\mu yx)}
    \ket{\bar{x},\bar{\mu}, \bar{a}}_{\text{sym}}
    \otimes \ket{\widetilde{B};y,\mu}_{\text{bdy}}.
\end{align}
Each state on the right hand side is precisely the Ishibashi state, and the coefficient is shown to satisfy the Cardy condition \cite{Choi:2024tri}. Although we don't need the explicit form of $\Psi$ for the purpose of the current work, interested readers may refer \cite{Choi:2024tri} for further details.

At last, we make the remark that the framework discussed in this section applies to the case where $\mathcal{C}$ is a fusion category, meaning that the symmetry is finite. However, in this work we also analyze the case of a $U(1)$ symmetry in a CFT. Fortunately, the $U(1)$ SymTFT admits a Lagrangian field theory description, which, to the best of our knowledge, suffices for our purposes.

\section{Criteria for Simple Symmetric Conformal Boundaries}\label{sec: criteria}

From the lore, we are expected to search for a simple conformal boundary condition that satisfies certain symmetry properties. In what follows, we clarify, within the SymTFT framework, the notions of simplicity, symmetry, and conformality of a boundary condition, respectively. We then synthesize these elements to formulate a criterion for the lore being hold.

\subsection{Simpleness Condition}

In a 2d boundary CFT, a boundary state $\ket{B}$ is simple if it can not be decomposed to be a sum of two boundary states with non-negative integer coefficients. This can also be phrased as no topological (vanishing scaling dimension) boundary-point point operators other than identity.

In the SymTFT framework, because $\ket{B}$ transforms under an \emph{irreducible} NIM-rep of the fusion algebra, the topological boundary $\mathcal{B}^{\text{bdy}}$ is simple. Moreover, we assume that the physical corner $\tilde{B}$ transforms in the regular module with respect to $\mathcal{B}^{\mathrm{bdy}}$, so that there is no topological junction between $\widetilde{B}$ and any non-trivial lines $\alpha\in \mathcal{B}^{\text{bdy}}$.\footnote{This also follows from the fact that $\tilde{B}$ can be thought of as a conformal boundary condition of the gauged CFT where we gauge the original CFT by the algebra object that corresponds to the module category of boundary conditions. We are grateful for Yichul Choi and Brandon Rayhaun for highlighting this point. 
}
Then picking a simple object in such regular module yields a simple $\widetilde{B}$, and picking a simple object in $(\mathcal{B}^{\text{sym}}, \mathcal{B}^{\text{bdy}})$-bimodule category $\underline{B}$ yields a simple $\underline{B}$.  
Then simpleness of $\ket{B}$ is equivalent to the simpleness of the symmetry corner $\underline{B}$ and the physical corner $\widetilde{B}$. We will impose the simpleness condition throughout the discussion below.

\subsection{Symmetric Condition}\label{sec: sym condition}

There are two notions for a boundary condition to be symmetric with respect to fusion category symmetry, known as being \emph{weakly} or \emph{strongly} symmetric. It has been proven \cite{Choi:2023xjw} that the two notions coincide in the case of invertible symmetry, while bifurcate in the case of non-invertible symmetry.   

For a fusion subcategory $\mathcal{D}\subset \mathcal{C}$,
a weakly $\mathcal{D}$-symmetric boundary condition allows a topological junction between any symmetry line $a\in\mathcal{D}$ and the boundary condition $B$, i.e. there exists an $a$-twisted boundary condition for any $a$ in $\mathcal{D}$. In the SymTFT framework, there exists a topological junction between $a$ and the symmetry corner $\underline{B}$ in the middle figure of Figure \ref{fig:symtft bdy state}.

An (untwisted) boundary condition $B$ is strongly $\mathcal{D}$-symmetric if the parallel fusion with $a\in\mathcal{D}$ gives $d_a$ copies of $B$, where $d_a$ is the quantum dimension of $a$. In the SymTFT framework, 
\begin{equation}
    a\otimes \underline{B} = d_a \underline{B}.
\end{equation}
Note that $\underline{B}$ as the interface is in the category of $\mathcal{D}$-module.

In summary, we arrive at a condition constraining the existence of $\mathcal{D}$-symmetric symmetry corner:

\paragraph{Symmetric condition:} A boundary condition $B$ is (strongly/weakly) $\mathcal{D}$-symmetric if we choose $\mathcal{B}^{\text{bdy}}$ such that the symmetry corner $\underline{B}$ between $\mathcal{B}^{\text{sym}}$ and  $\mathcal{B}^{\text{bdy}}$ is (strongly/weakly) $\mathcal{D}$-symmetric. 
\\~

Note that the \textbf{symmetric condition} constrains the choice of $\mathcal{B}^{\text{bdy}}$.

\subsection{Conformal Condition}

We further clarify a necessary condition for the boundary state $\ket{B}_{a,z}$ to be conformal in SymTFT set-up.

We start with the closed-string channel. There are two ways to decompose the Hilbert space of a closed CFT: one by the global symmetry, and the other by the Virasoro algebra. The SymTFT is tailored to the first case, in which the Hilbert space and the boundary state are decomposed as in \eqref{hilb space decompose} and \eqref{bdy state}, with respect to the symmetry $\mathcal{C}$. 

Alternatively, the $a$-twisted Hilbert space can be decomposed into a direct sum of representations of the chiral and anti-chiral Virasoro algebras:
\begin{align}\label{H^a decomplose}
    \mathcal{H}^a = \bigoplus_{i, j} M^a_{i j} \, V_i \otimes \bar{V}_j.
\end{align}
Since in field theory the symmetry operators commute with the stress tensor $T^{\mu\nu}$, and hence with the Virasoro algebras, we conclude that each sub-Hilbert space, namely the multiplet $\mathcal{H}^a_\mu \subset \mathcal{H}^a$ associated with $\mathcal{C}$, admits a similar decomposition:
\begin{align}\label{H^a_mu decomplose}
    \mathcal{H}^a_\mu = \bigoplus_{i, j \in I_\mu} M^{a}_{\mu,i j} \, V_i \otimes \bar{V}_j,
\end{align}
where $M^{a}_{\mu,i j}\in\mathbb{Z}_{\geq 0}$ determines the conformal families in space $\mathcal{H}^a_\mu$.

A boundary state is conformal if it preserves half of the Virasoro symmetry, 
\begin{equation}\label{eq:conformalcondition}
    (L_n - \bar{L}_{-n}) \ket{B}_{a,z} =0.
\end{equation}
This equation has been completely solved in vector spaces of the form \eqref{H^a decomplose} or \eqref{H^a_mu decomplose} by Ishibashi \cite{Ishibashi:1988kg, Onogi:1988qk}, which is reviewed in Appendix \ref{ishibashi}. The conclusion is, that all solutions are spanned by Ishibashi states. In particular, Ishibashi states exist only in the subspace $V_i \otimes \bar{V}_j$ satisfying $h_i = h_j$\footnote{We emphasis that Ishibashi's result is a constraint on conformal family $V_i\otimes \bar{V}_j$, not a particular state, since $(h_i,h_j)$ is the conformal weights of the primary operator in the conformal family.}, and it is unique once normalization is fixed.

In order for the basis in \eqref{bdy state}, 
\begin{equation}
    \ket{\bar{x},\bar{\mu}, \bar{a}}_{\text{sym}}  \otimes\ket{\widetilde{B};y,\mu}_{\text{bdy}}\in W^a_\mu \otimes \mathcal{V}_\mu=\mathcal{H}^a_\mu=\bigoplus_{i j} M^a_{\mu,i j} V_i \otimes \bar{V}_j,
\end{equation}
to satisfy the conformal condition \eqref{eq:conformalcondition}, there must exist a spin 0, that is, $h_i=h_j$, conformal family $V_i \otimes \bar{V}_j$ within $\mathcal{H}^a_\mu$. Note that the subspace $\mathcal{H}^a_\mu$ is determined by $(a,\mu)$, where $\mu\prec \mathcal{A}^{\text{bdy}}$ with $F_{\mathcal{B}^{\text{sym}}}(\mu)\succ a$ for a given $a$.

We point it out that the Lorentz spin of a primary operator in $\mathcal{H}^a_\mu$ equals the topological spin of $\mu$ plus a integer. The condition here requires that the spin of primary operator is strictly zero, not a non-zero integer. 

Since $\ket{\bar{x},\bar{\mu}, \bar{a}}_{\text{sym}}  $ is topological, conformal condition is imposed only on $\ket{\widetilde{B};y,\mu}_{\text{bdy}}$ which comes from the non-topological physical corner and is independent of the choice of $a$ in $ \mathcal{B}^{\text{sym}}$. Hence to obtain the full conformal condition on the physical corner,  one should union the above spin 0 condition for all possible $a$ that are terminable on $\underline{B}$. In terms of the rightmost figure in Figure \ref{fig:symtft bdy state}, union over $a$ means one requires spin 0 condition for all possible $\mu\prec\mathcal{A}^{\text{bdy}}$.

If the spin 0 condition is violated, we claim that the physical corner $\widetilde{B}$ cannot be simple, which is proved below. If some $\mu\prec\mathcal{A}^{\text{bdy}}$ does not satisfy the condition above, then the only solution to \eqref{eq:conformalcondition} in the subspace $\mathcal{H}^a_\mu=W^a_\mu \otimes \mathcal{V}_\mu$ is zero, indicating $\ket{\widetilde{B};y,\mu}_{\text{bdy}}=0$. However, the regularized overlap of the half-Ishibashi state is, as calculated in \cite{Choi:2024tri}
\begin{align}
    {\vphantom{\ket{\widetilde{B}; y,\mu}_{\text{bdy}}}}_{\text{bdy}}\bra{\widetilde{B}; y, \mu} \tilde{q}^{\frac12(L_0+ \bar{L}_0 - \frac{c}{12})} \ket{\widetilde{B}; y,\mu}_{\text{bdy}}=\frac{1}{\sqrt{S_{11}}}\sum_{\alpha\in\mathcal{B}^{\text{bdy}}}{}^{\mathcal{B}^{\text{bdy}}\mathcal{B}^{\text{bdy}}}\Psi^{11}_{\al\al(\mu yy)}\mathbf{Z}_\al(\delta),\quad \widetilde{q}=e^{-4\pi/\delta},
\end{align}
where we sum over simple objects $\al\in \mathcal{B}^{\text{bdy}}$ and $\mathbf{Z}_\al$ is the representation basis partition function defined in \cite{Choi:2024tri}. Note that the coefficient in the right-hand side is proportional to the half-braiding matrix associated with $\mathcal{Z}(\mathcal{B}^{\text{bdy}})$ by definition, thus the coefficient of $\mathbf{Z}_1$ is not zero. Since, by assumption, the left-hand side can only be zero, we arrive at a contradiction if there are no other nontrivial $\al$ labeling boundary-changing operators with vanishing scaling dimension. 

In summary, we arrive at a  necessary condition constraining the existence of simple conformal physical corner:

\paragraph{Conformal condition:} A necessary condition for the physical corner between $\mathcal{B}^{\text{bdy}}$ and $\widetilde{Q}$ to be conformal and simple is that, for each $\mu$ condensable on $\mathcal{B}^{\text{bdy}}$, there exists a spin-0, i.e. $h_i=h_j$, conformal family $V_i \otimes \bar{V}_j$ in the subspace $\mathcal{H}^a_\mu$ for some $a$ in $\mathcal{B}^{\text{sym}}$.
\bigskip

Note that the conformal condition above is also a constraint on the choice of $\mathcal{B}^{\text{bdy}}$.

\subsection{Criteria}\label{sec: criteriasub}

Now we put together all the conditions, and propose criteria for the existence of simple, symmetric, and conformal boundary condition of a CFT in SymTFT framework. 

Since we limit our discussion on \emph{conformal} boundary conditions, the idea is to enumerate possible $\mathcal{B}^{\text{bdy}}$'s satisfying the conformal condition above, and then check the symmetry properties by examining symmetry corners.

More concretely, we propose the following steps:
\begin{enumerate}
    \item Classify all topological boundary conditions $\{\mathcal{B}\}$ of the 3d SymTFT.
    \item For any $\mathcal{B}$, check the \textbf{conformal condition}, i.e. for \emph{each} $\mu$ condensable on $\mathcal{B}$, we need to find at least one spin-0 conformal family in $\mathcal{H}_\mu^b$ for some $b\in \mathcal{C}$. The condition being satisfied is a necessary condition for $\mathcal{B}$ serving as $\mathcal{B}^{\text{bdy}}$ to realize a conformal boundary state for CFT.
    \item Determining the simple symmetry corners for each choice of $\mathcal{B}^{\text{bdy}}$, i.e. simple objects in the $(\mathcal{B}^{\text{sym}}, \mathcal{B})$-bimodule category.
\end{enumerate}

Since the symmetry property of a conformal boundary state is governed by the symmetry corner in the SymTFT framework as discussed in Section \ref{sec: sym condition}, going through the 3-step procedure, we actually determine all possible symmetry multiplets that can be formed by simple conformal boundary states. 

To see when the lore introduced in the Introduction section can be violated, we need to further impose the \textbf{symmetric condition} in step 3.

The anomaly is defined as the obstruction to gauging with a Frobenius algebras $A$ in $\mathcal{D}$. \footnote{As reviewed in the Introduction, two notions of anomaly free conditions are introduced in \cite{Choi:2023xjw}---strongly and weakly anomaly free. They differ by whether the Frobenius algebra is maximal.} Start with $\mathcal{B}^{\text{sym}}$ and gauge the Frobenius algebra $A\in\mathcal{D}$. We obtain a topological boundary $\mathcal{B}$ of SymTFT. By choosing $\mathcal{B}^{\text{bdy}}=\mathcal{B}$, the symmetry corner $\underline{B}$ is the half-gauging interface, which is an object in $\mathcal{C}_A$, i.e. the category of right $A$-modules in $\mathcal{C}$ (also identified as the category of left $\mathcal{C}$-module). 

Due to the result in the section 2 of \cite{Diatlyk:2023fwf}, the (strong or weak) \textbf{symmetric condition} can always be satisfied by this $\mathcal{B}$ and some simple object in $\mathcal{C}_A$. 

More concretely, every $a$ in $A$ admits topological junctions on $\underline{B}\in\mathcal{C}_A$ via the non-empty $\operatorname{Hom}$ space:
\begin{align}\label{eq:Hom}
    \operatorname{Hom}_{\mathcal{C}_A}(a\otimes\underline{B}, \underline{B})=\operatorname{Hom}_{\mathcal{C}}(a, \underline{\operatorname{Hom}}(\underline{B},\underline{B}))=\operatorname{Hom}_{\mathcal{C}}(a,A)\neq\emptyset,
\end{align}
which ensures that the symmetry corner $\underline{B}$ is weakly symmetric. Furthermore, if $\mathcal{D}$ is strongly anomaly-free, let $A$ be the maximal haploid (or connected) Frobenius algebra in the Morita equivalent class. The algebra object $A$, as the regular right $A$-module, is a simple object in $\mathcal{C}_A$.  The simpleness follows from
\begin{align}
    \operatorname{Hom}_{\mathcal{C}_A}(A,A)=\operatorname{Hom}_{\mathcal{C}}(1,A)\simeq\mathbb{C}\quad\text{(haploid)}.
\end{align}
Hence, by choosing $\underline{B}=A\in\mathcal{C}_A$ with $A= \oplus_{a\in \mathcal{D}} d_a a$, we find from the same computation in \eqref{eq:Hom}
\begin{eqnarray}
	\operatorname{Hom}_{\mathcal{C}_A}(a\otimes A, A)=\operatorname{Hom}_{\mathcal{C}}(a,A) = d_a,
\end{eqnarray}
which guarantees $a\otimes \underline{B} = d_a \underline{B}$. Therefore we obtain a strongly symmetric symmetry corner with respect to the strongly anomaly free symmetry $\mathcal{D}$.

However, demanding $\mathcal{B}$ to further satisfy the \textbf{conformal condition} may have obstructions, depending on the dynamical data of the CFT. When such $\mathcal{B}$ exists, the lore is obeyed, and appeared in many examples \cite{Li:2022drc}. When such $\mathcal{B}$ does not exist, the lore is violated. This is the main mechanism of how the lore can be violated, and we will illustrate this with several examples in the following sections.

Since the \textbf{symmetric condition} can always be satisfied by a topological boundary $\mathcal{B}$ as discussed above,  despite the \textbf{conformal condition} can not be satisfied by the same $\mathcal{B}$ in every CFT, we assume that it can be satisfied in some CFT. So we hypothesize the following improved lore: 
\paragraph{Improved Lore:} Given symmetry $\mathcal{C}$ and its (strongly or weakly) anomaly-free subcategory $\mathcal{D}\subset\mathcal{C}$, there exists a CFT which has a simple, (strongly or weakly)  $\mathcal{D}$-symmetric, conformal boundary.\footnote{We assume that any symmetry $\mathcal{C}$ can be realized in CFTs with one vacuum.}

In fact, the improved lore follows the assumption that any symmetry $\mathcal{C}$ can be realized in CFTs with one vacuum.\footnote{We are grateful to Brandon Rayhaun for sharing the proof.} The construction was discussed in \cite[Section 4.1]{Choi:2023xjw}. Suppose a single-vacuum CFT $Q$ has $\mathcal{C}$ symmetry, which contains $\mathcal{D}$ as a strongly anomaly-free subcategory. One can gauge $\mathcal{D}$ on half space and the gauging interface is described by a $\mathcal{D}$ module category with a single object (thanks to strongly anomaly free condition). One then folds the theory along the gauging interface and obtains $Q\otimes \overline{Q/\mathcal{D}}$ with $\mathcal{D}$-preserving boundary condition, where $\mathcal{D}$ only acts on the first component of the product theory. In the case of $\mathcal{D}$ being weakly symmetric, one can pick the gauging interface to be described by a simple object in the $\mathcal{D}$ module category on which lines in $\mathcal{D}$ can topologically terminate, and the rest of the discussions follow.

\section{The Lore Violated: $\mathbb{Z}_p^m\times\mathbb{Z}_q^w$ in Compact Boson}\label{sec: boson}

In this section, we focus on the finite subgroup of group symmetry in 2d CFT. This example was mentioned in \cite{SMGtalk} by Philip Boyle Smith in a talk in SCGP workshop, in which he mentioned that this example was noticed by Masataka Watanabe.\footnote{Yunqin Zheng is also in debt to Philip Boyle Smith for making him aware of this example during an IPMU tea time, and is grateful to Yichul Choi for a related discussion at the SCGP workshop on Symmetric Mass Generation.} By applying the criteria proposed in Section \ref{sec: criteria}, we see the lore is violated. 

Consider the 2d $c=1$ compact boson CFT, with $U(1)^m\times U(1)^w$ symmetry. The $U(1)^m$ momentum symmetry shifts the $2\pi$ periodic scalar field $\f$ by a constant. The $U(1)^w$ winding symmetry shifts the T-dual field $\widetilde{\f}$ by a constant. As is well known that the $U(1)^m\times U(1)^w$ symmetry has a mixed anomaly, which can be derived by performing a gauge transformation to the action coupling to background fields:
\begin{align}\label{2d boson action}
    S[A, \widetilde{A}]=\int_{M^2} d\tau dx \left(\frac{R^2}{4\pi}(\partial_\mu\f-A_\mu)^2+\frac{i}{2\pi}\epsilon^{\mu\nu}\widetilde{A_\mu}(\partial_\nu\f-A_\nu)\right), 
\end{align}
where $A$ and $\widetilde{A}$ are background $U(1)$ gauge fields. The 3d anomaly inflow is given by 
\begin{align}
    \frac{i}{2\pi}Ad\widetilde{A}.
\end{align}

Now we consider a finite subgroup $\mathbb{Z}^m_p\times\mathbb{Z}^w_q\subset U(1)^m\times U(1)^w$ with $\gcd(p,q)=1$. The subgroup is anomaly-free. One way to see this is through the cohomology classification $H^3(\mathbb{Z}_p\times\mathbb{Z}_q, U(1))=\mathbb{Z}_p\times \mathbb{Z}_q\times \mathbb{Z}_{\gcd(p,q)}$, where the first two factors are self anomalies of $\bZ_p$ and $\bZ_q$ respectively, which we know to vanish. The last factor comes from the mixed anomaly, which is the trivial group $\mathbb{Z}_1$ in our case. A more concrete way is to show that the anomaly inflow becomes trivial after restricting to the subgroup. We set $A=\frac{2\pi}{p}c$ and $\widetilde{A}=\frac{2\pi}{q}\widetilde{c}$, where $c\in H^1(M^2, \mathbb{Z}_p)$ and $\widetilde{c}\in H^1(M^2, \mathbb{Z}_q)$. Since the gauge fields are flat, we have
\begin{align}
    \int\delta c=0\mod p\quad\text{and}\quad\int\delta \widetilde{c}=0\mod q.
\end{align}
By the Chinese Remainder theorem, for $\gcd(p,q)=1$, there exists $x,y\in\mathbb{Z}$ such that $px+qy=1$, we can rewrite the anomaly inflow as
\begin{align}
    \frac{i}{2\pi}Ad\widetilde{A}=\frac{2\pi i}{pq}c\delta\widetilde{c}=2\pi i\frac{px+qy}{pq}c\delta\widetilde{c}=2\pi i \left(xc\frac{\delta\widetilde{c}}{q}+y\frac{\delta c}{p}\widetilde{c}\right)+\delta\left(2\pi i\frac{c\cup\widetilde{c}}{pq}\right).
\end{align}
The first two terms take value in $2\pi i\mathbb{Z}$ after taking the integral, and the last term can be removed by adding a local counterterm $\frac{i}{2\pi}A\we\widetilde{A}$ in the action \eqref{2d boson action}. Thus, we confirm that the finite subgroup $\mathbb{Z}^m_p\times\mathbb{Z}^w_q$ is anomaly-free.

\subsection{Space of Conformal Boundary Conditions}

So far we have just discussed the symmetry properties; now let us turn to the conformal boundary state. According to the lore, for any anomaly-free subgroup, there is supposed to exist conformal boundary states preserving the symmetry. The simple conformal boundary conditions have been studied in \cite{Choi:2025ebk,Gaberdiel:2001zq, Gaberdiel:2001xm}, which we summarize below.

When the radius $R=u/v$ is rational with $\gcd(u,v)=1$, the space of simple conformal boundary conditions is
\begin{eqnarray}\label{eq:Ruvbdyspace}
				R=\frac{u}{v}: \qquad S^1_{\text{Dir}} \cup S^1_{\text{Neu}} \cup ~ \frac{SU(2)'}{\bZ_u\times \bZ_v}. 
\end{eqnarray}
The first two factors are Dirichlet boundary conditions and Neumann boundary conditions respectively, which are known to form $S^1$'s. The last factor is a coset (with special points deleted). Here $SU(2)/(\bZ_u \times \bZ_v)$ is parameterized by
\begin{eqnarray}\label{eq:abab}
		\left| 
		\begin{pmatrix}
			a & b\\
			-b^* & a^*
		\end{pmatrix}
		\right\rangle_{R=\frac{u}{v}}, \quad |a|^2 + |b|^2 =1, \quad a\sim a e^{2\pi i/v}, \quad  b\sim b e^{2\pi i/u} 
\end{eqnarray}
and the prime means specializing to the space with $ab\neq 0$. The last set of boundary conditions descend from the $SU(2)$ boundary conformal manifold via gauging $\bZ_v^{m}\times \bZ_u^w$. At the self-dual radius $R=1$, the three branches combine to form the $SU(2)$ group manifold. 

When the radius $R$ is irrational, the space of simple conformal boundary conditions is
\begin{eqnarray}
    R=\text{irrational}: \qquad S^1_{\text{Dir}} \cup S^1_{\text{Neu}}.
\end{eqnarray}
Those correspond the coset factor in \eqref{eq:Ruvbdyspace} become non-compact, and was discussed in \cite{Janik:2001hb, Gaberdiel:2001zq}. We restrict ourselves to the compact case because it is difficult (if not impossible) to define the notion of a simple boundary state in the non-compact case, therefore we only focus on the two $S^1$ factors, i.e. the Dirichlet and Neumann boundaries. 

\subsection{Symmetries Preserved by the Conformal Boundary Conditions}

What symmetries are preserved by these boundary conditions? It is well known that the Dirichlet boundaries preserve the $U(1)$ winding symmetry but break the $U(1)$ momentum symmetry, whereas the Neumann boundaries preserve the $U(1)$ momentum symmetry but break the $U(1)$ winding symmetry.

To determine the symmetry preserved by the boundary condition in the coset factor of \eqref{eq:Ruvbdyspace}, we need to determine how the $U(1)^m \times U(1)^w$ acts on the parameters $a,b$. The strategy is to identify the symmetry action in the $R=1$ case, and then identify the action in the $R=u/v$ case via gauging $\bZ_v^m \times \bZ_u^w$.

We start with $R=1$, where the $U(1)^m \times U(1)^w$ acts via 
\begin{eqnarray}
	\begin{split}
		U(1)^m: &  \quad a\to a e^{i \alpha}, \quad b\to b,\\
		U(1)^w: & \quad a \to a, \quad b\to b e^{i \beta}.
	\end{split}
\end{eqnarray}
The theory with radius $R=\frac{u}{v}$ can be obtained from $R=1$ theory by gauging the $\bZ_v^m\times \bZ_u^w$ subgroup. 
Note that $a$ transforms the same way as $e^{i \phi}$ in the $R=1$ theory, where $\phi$ is the compact scalar, and $b$ transforms the same way as $e^{i\theta}$ where $\theta$ is the dual compact scalar. Combining with the fact that the compact boson $\phi'$ in $R=\frac{u}{v}$ theory is related to $\phi$ of the $R=1$ theory by $\phi'= \frac{v}{u}\phi$, we find that under $U(1)^m$ transformation $\phi'\to \phi'+ \alpha'$ implies $\phi\to \phi+ \frac{u}{v}\alpha'$, from which we see how $U(1)^m$ acts on the boundary state. One can similarly work on how $U(1)^w$ acts by T-duality. The results are
\begin{eqnarray}\label{eq:U1transformations}
	\begin{split}
		U(1)^m: &\quad a\to a e^{i u \alpha'/v }, \quad b\to b,\\
		U(1)^w: &\quad a\to a, \quad b\to b e^{i v \beta'/u}.
	\end{split}
\end{eqnarray}
By combining with the identification relation in \eqref{eq:Ruvbdyspace}, only the $\bZ_u^m\times \bZ_v^w$ subgroup of $U(1)^m \times U(1)^w$ preserved.

As summarized in Table \ref{tab:R=u/v} and \ref{tab:Rirrational}, there is no known simple conformal boundary condition preserving $\bZ_p^m\times \bZ_q^w$, unless $R=\frac{p}{q}$.

\begin{table}[t]
    \centering
    \begin{tabular}{|c |c| c| c|}
				\hline
				\text{Boundary Conditions}	& $U(1)^m$ & $U(1)^w$ & $\bZ_p^m \times \bZ_q^w$\\
				\hline
				$S^1_{\text{Dir}}$ & $\times$ & $\checkmark$ & $\times$\\
				$S^1_{\text{Neu}}$ & $\checkmark$ & $\times$ & $\times$\\
				${SU(2)'}/({\bZ_u\times \bZ_v})$ & $\times$ & $\times$ & $\checkmark$  iff $\frac{u}{v}= \frac{p}{q}$\\
				\hline
			\end{tabular}
    \caption{Known conformal boundary conditions and the preserved symmetries for rational radius $R=u/v$. }
    \label{tab:R=u/v}
\end{table}

\begin{table}[t]
    \centering
    \begin{tabular}{|c |c| c| c|}
				\hline
				\text{Boundary Conditions}	& $U(1)^m$ & $U(1)^w$ & $\bZ_p^m \times \bZ_q^w$\\
				\hline
				$S^1_{\text{Dir}}$ & $\times$ & $\checkmark$ & $\times$\\
				$S^1_{\text{Neu}}$ & $\checkmark$ & $\times$ & $\times$\\
				\hline
			\end{tabular}
    \caption{Known conformal boundary conditions and the preserved symmetries for irrational radius $R$. }
    \label{tab:Rirrational}
\end{table}

\subsection{SymTFT, Topological Boundaries, and Physical Boundary}\label{sec: symtftbosonfield}

In the following two sections, We use our criteria to show that at a generic radius, the anomaly-free-$\mathbb{Z}_p^m\times\mathbb{Z}_q^w$ symmetry cannot be preserved by a simple conformal boundary state.

We start our discussion from the $U(1)\times U(1)$ SymTFT \cite{Antinucci:2024zjp, Brennan:2024fgj, Apruzzi:2024htg}, and realizing it's topological boundaries. We do the field theory version here, and leave the Euclidean lattice version to the appendix \ref{lattice}.

The continuous SymTFT of a 2d theory with $U(1)$ symmetry is described by a 3d BF theory: $\frac{i}{2\pi}\omega \, da$, where $\omega$ is a 1-form $\mathbb{R}$-valued gauge field, and $a$ is a 1-form $U(1)$ gauge field. For $U(1)\times U(1)$ symmetry with a mixed anomaly, the action for the SymTFT is
\begin{align}
    \frac{i}{2\pi} \int_{M^3} \omega da + \widetilde{\omega} d\widetilde{a} + ad\widetilde{a},
\end{align}
where $\omega, \widetilde{\omega}$ are $\mathbb{R}$-valued 1-form gauge fields, and $a, \widetilde{a}$ are $U(1)$-valued 1-form gauge fields. By performing a field redefinition: $b = a+\widetilde{\omega}, \widetilde{b}=\widetilde{a}+\omega$, the action is simplified to 
\begin{align}
    \frac{i}{2\pi} \int_{M^3} b d\widetilde{b} - \frac{i}{2\pi} \int_{M_3} \widetilde{\omega}d\omega.
\end{align}
The first term gives raise to a trivial TQFT \cite{Witten:2003ya, Hsin:2016blu}, so we focus on the remaining nontrivial term:
\begin{align}\label{S3}
    S_3 = -\frac{i}{2\pi} \int_{M^3} \widetilde{\omega}d\omega,
\end{align}
where $\omega,\widetilde{\omega}$ are $\mathbb{R}$-valued 1-form gauge fields, transforming as $\omega\to \omega+ d\eta$, $\widetilde{\om} \to \widetilde{\om}+d\widetilde{\eta}$, with $\eta, \widetilde{\eta}$ valued in $\mathbb{R}$. Equation \eqref{S3} defines the bulk action of the 3d SymTFT. The theory admits line operators labeled by pairs of $\mathbb{R}$-valued numbers: $L_{(x,y)} \coloneqq \exp\left(i x \int \omega_1 + i y\int \omega_2\right)$,
with braiding  given by
\begin{align}
    \left\langle \exp(ix\int_{\gamma}\omega)\exp(iy\int_{\widetilde{\gamma}}\widetilde{\om})\right\rangle = e^{2\pi i x y \#(\gamma, \widetilde{\gamma})}.
\end{align}
Here $\#(\cdot,\cdot)$ denotes the linking number. 
Thus, the topological spin of $L_{(x,y)}$ is $\exp(2\pi ixy)$.

Let us discuss the topological boundary conditions of this 3d TQFT. The action \eqref{S3} is not gauge invariant on a manifold with boundary, but with a boundary term
\begin{align}
    \frac{i}{2\pi}\int_{\partial M^3} d\widetilde{\eta}\we\om.
\end{align}
An appropriate boundary condition must cancel this variation to ensure gauge invariance. There are two natural choices:
\begin{align}
\begin{split}
    \text{Dirichlet:}\quad &\om\big|_{\partial M^3}=0,\\
    \text{Neumann:}\quad &\widetilde{\om}\big|_{\partial M^3}=0.
\end{split}
\end{align}
Alternatively, one can specify a boundary condition by identifying the corresponding Lagrangian algebra\footnote{Strictly speaking, this notion is not mathematically rigorous for continuous SymTFTs, since the theory contains infinitely many simple lines.}, that is, the set of bulk lines that can condense on the boundary. We summarize them in the third column of Table \ref{lag alg}.

Searching for all Lagrangian algebras of \eqref{S3}, we have an additional topological boundary condition, where lines of the form $L_{(\frac{n}{r}, mr)}$ can condense, with $r\in \mathbb{R}$ and $n,m\in \bZ$.  It can be realized by introducing on the boundary a Lagrange multiplier $\theta$, a $2\pi$-periodic scalar field transforming as $\theta \to \theta + \eta$. The boundary action is
\begin{align}\label{S2sym}
    S_2=\frac{i}{2\pi}\int_{\partial M^3}\widetilde{\om}(rd\theta-\om).
\end{align}
One can check that the total action $S_3 + S_2$ is gauge invariant. Varying with respect to $\widetilde{\omega}$ enforces $\omega$ to have $2\pi r\mathbb{Z}$ holonomy on the boundary. Conversely, summing over all winding sectors of $\theta$ constrains $\widetilde{\omega}$ to have $\frac{2\pi}{r}\mathbb{Z}$ holonomy. This is precisely the condition under which lines $L_{(\frac{n}{r}, mr)}$ can condense.

We can think of the topological boundary conditions for TQFT \eqref{S3} being parameterized by a continuous real parameter $r\geq0$, denoted as $\mathcal{B}_r$. The Dirichlet boundary corresponds to the $r=0$ limit, and the Neumann boundary corresponds to the $r\to\infty$ limit. 

We choose $r=1$ case, then nontrivial lines support on the third boundary condition are labeled by two $2\pi$-periodic variable: 
\begin{align}
    L_{(\frac{\al}{2\pi}, \frac{\beta}{2\pi})}=\exp\left(\frac{i\al}{2\pi}\int\om+\frac{i\beta}{2\pi}\int\widetilde{\om}\right), \quad \al\sim\al+2\pi, \quad \beta\sim\beta+2\pi.
\end{align}
This boundary serves as our symmetry boundary in the SymTFT construction for compact boson, $\mathcal{B}^{\text{sym}}\coloneqq\mathcal{B}_1$. The momentum and winding symmetry operators, supporting on the symmetry boundary,  are given by 
\begin{align}
    U^m_\beta\coloneqq\exp(i\beta J^m)=L_{(0, \frac{\beta}{2\pi})}, \quad U^w_\al\coloneqq\exp(i\al J^w)=L_{(\frac{\al}{2\pi},0)}.
\end{align}
The bulk-to-symmetry-boundary forgetful functor is
\begin{align}
    F_{\mathcal{B}_1}(L_{(x, y)})=L_{(\frac{\al}{2\pi}, \frac{\beta}{2\pi})},
\end{align}
where $x\in \mathbb{Z}+\frac{\al}{2\pi}$ and $y\in\mathbb{Z}+\frac{\beta}{2\pi}$.

\begin{table}
\centering
\begin{tabular}{|c|c|c|}
\hline
Notations &Boundary conditions & Lagrangian algebras \\
\hline
$\displaystyle\mathcal{B}_{0}$ & $\displaystyle\om\big|_{\partial M^3}=0$ & $\displaystyle\bigoplus_{x\in\mathbb{R}}L_{(x,0)}$ \\[6pt]
\hline
$\displaystyle\mathcal{B}_{\infty}$& $\displaystyle\widetilde{\om}\big|_{\partial M^3}=0$ & $\displaystyle\bigoplus_{y\in\mathbb{R}}L_{(0,y)}$ \\[6pt]
\hline
$\displaystyle\mathcal{B}_{r}$& $\displaystyle S_2=\frac{i}{2\pi}\int_{\partial M^3}\widetilde{\om}(rd\theta-\om)$ & $\displaystyle \bigoplus_{n,m\in\mathbb{Z}}L_{(\frac{n}{r}, mr)}$ \\[6pt]
\hline
\end{tabular}
\caption{Gapped simple topological boundary conditions}\label{lag alg}
\end{table}

In the same spirit we give the physical boundary $\widetilde{Q}$ a Lagrangian description:
\begin{align}\label{S2phy}
    S_2^{\text{phy}}=\frac{R^2}{4\pi}\int_{M^2} (dX-\om)\star(dX-\om).
\end{align}
Here the field $X$ is a real scalar, not compact. After shrinking the sandwich with $\mathcal{B}^{\text{sym}}=\mathcal{B}_1$, both $\widetilde{\om}$ and $\om$ are integrated out, and we are left with a purely 2d theory with action
\begin{align}
    \frac{R^2}{4\pi}\int_{M^2}(dX-d\theta)\star(dX-d\theta).
\end{align}
Remember that $\theta$ is $2\pi$ periodic, so we define a $2\pi$ periodic scalar $\f=\theta-X$, then the action can be rewritten as 
\begin{align}
    \frac{R^2}{4\pi}\int_{M^2}d\f\star d\f,
\end{align}
which reproduces compact boson at radius $R$. The action \eqref{S3}, \eqref{S2sym} and \eqref{S2phy} together give a complete description of the SymTFT for compact boson.\footnote{More generally, if we take $\mathcal{B}^{\text{sym}}= \mathcal{B}_r$, then shrinking the sandwich gives $\frac{R^2}{4\pi}\int_{M^2} (dX-rd\theta)\star (dX-rd\theta)$. Introducing the $2\pi$ periodic scalar $\phi= \theta- X/r$ simplifies the action to $\frac{R^2}{4\pi r^2}\int_{M^2} d\phi \star d\phi$. Note that the radius of the compact boson is now $R/r$. This vividly shows how gauging $U(1)$ symmetry is related to the radius.}

Finally, let us spell out the operators in the physical junction space $\mathcal{V}_\mu$. For a (generally nonlocal) operator $O_{x,y} = \exp(ix\f+iy\widetilde{\f})$ in the compact boson theory, its conformal family resides in the space $\mathcal{V}_\mu$, where $\mu = L_{(x,y)}$.  
The operator $O_{x,y}$ belongs to the $F_{\mathcal{B}_1}(L_{(x,y)}) = L_{\left(\frac{\alpha}{2\pi}, \frac{\beta}{2\pi}\right)}=U^m_\beta U^w_\al$-twisted Hilbert space.

\subsection{Criteria Applied: No $\bZ_p^m\times \bZ_q^w$ Preserving Boundary at $R\neq \frac{p}{q}$}

Let us follow the procedure in Section \ref{sec: criteriasub} to study the simple symmetric conformal boundary conditions of the compact boson.

Step 1 has already been done in the last section: all topological boundaries of SymTFT are given by $\mathcal{B}_r$'s. For step 2, the conformal weight of the operator $O_{x,y} = \exp(ix\f+iy\widetilde{\f})$ is 
\begin{align}
    h=\frac{1}{4}\left(\frac{x}{R}+yR\right)^2, \quad \bar{h}=\frac{1}{4}\left(\frac{x}{R}-yR\right)^2, 
\end{align}
meaning that the spin is given by $s=h-\bar{h}=xy$.

For the topological boundary $\mathcal{B}_0$, condensable lines $\mu=L_{(x,0)}$'s. The $U(1)$-conformal family in the subspace $\mathcal{H}^a_\mu$ is generated by $O_{x,0}$, which is spin-0, meaning that the boundary $\mathcal{B}_0$ satisfies the \textbf{conformal condition}. 

For the topological boundary $\mathcal{B}_\infty$, condensable lines are $\mu=L_{(0,y)}$'s. The $U(1)$-conformal family in the subspace $\mathcal{H}^a_\mu$ is generated by $O_{0,y}$, which is also spin-0, so that the boundary $\mathcal{B}_\infty$ satisfies the \textbf{conformal condition}. 

However, for $r>0$, the  boundary $\mathcal{B}_r$, the condensable lines are $\mu=L_{(\frac{n}{r},mr)}$'s. The $U(1)$-conformal family in subspace $\mathcal{H}^a_\mu$ are generated by $O_{\frac{n}{r},mr}$, with spin $s=nm\neq0$. But this does not immediately imply that the boundary $\mathcal{B}_r$ violates the \textbf{conformal condition}. However, it is known \cite{Gaberdiel:2001zq} that when both $h,\bar{h}$ are not perfect squares of half-integers (i.e. $\frac{n}{rR}\pm mrR \notin \bZ$), $O_{\frac{n}{r}, mr}$ is not only a $U(1)$-primary but also a \emph{Virasoro} primary. In this case, the non-zero spin $s\neq 0$ implies the violation of \textbf{conformal condition}. 

To compare with the known Dirichlet and Neumann boundary conditions, and answer the $\mathbb{Z}_p\times\mathbb{Z}_q$-preserving boundary state puzzle proposed at the beginning of this section, we further work out the third step in the criteria. Identifying a $\mathcal{B}^{\text{bdy}}$ boundary with certain symmetry multiplet can be done by analyzing the symmetry corner $\underline{B}$.

For the topological corner between $\mathcal{B}^{\text{sym}}=\mathcal{B}_1$ and $\mathcal{B}^{\text{bdy}}=\mathcal{B}_0$, we have the constraint that $\om=0$ and the holonomy of $\widetilde{\om}$ takes values in $2\pi\mathbb{Z}$. Thus, the only nontrivial line operator supporting on the corner is $L_{(0,\frac{\beta}{2\pi})}$, with $\beta\sim\beta+2\pi$. In other words, the symmetry corner is one-to-one correspond to the line: $\underline{B}_\beta=[L_{(0,\frac{\beta}{2\pi})}]$, labeled by a $S^1$ valued parameter.

Now we work out its behavior under the symmetry action. The parallel fusion with a winding symmetry line $U^w_\al=L_{(\frac{\al}{2\pi},0)}$ from $\mathcal{B}^{\text{sym}}$ is trivial:
\begin{align}
    L_{(\frac{\al}{2\pi},0)} \otimes \underline{B}_\beta=\underline{B}_\beta,
\end{align}
since $\om|_\text{corner}=0$. While the parallel fusion with a momentum symmetry line $U^m_{\beta'}=L_{(0,\frac{\beta'}{2\pi})}$ from $\mathcal{B}^{\text{sym}}$ shifts $\beta$:
\begin{align}
    L_{(0,\frac{\beta'}{2\pi})} \otimes \underline{B}_\beta=\underline{B}_{\beta+\beta'}.
\end{align}
Thus, the $\mathcal{B}^{\text{bdy}}=\mathcal{B}_0$ boundary realizes the $U(1)^m$-breaking-$U(1)^w$-preserving boundary states multiplet. We conclude that this construction realizes the Dirichlet boundary states.\footnote{The above discussion showed that setting $\mathcal{B}^{\text{bdy}}=\mathcal{B}_0$ strongly preserves $U(1)^{w}$. Alternatively, we can also see that $\mathcal{B}^{\text{bdy}}=\mathcal{B}_0$ also weakly preserves $U(1)^{w}$ by seeing that $L_{(x,0)}\prec \mathcal{A}^{\text{bdy}}=\mathcal{A}_0$ have a topological junction with the $U(1)^w$ generator $L_{(\frac{\alpha}{2\pi},0)}$ on the symmetry boundary. Indeed, it is shown in \cite{Choi:2023xjw} that for invertible symmetries strongly and weakly symmetric conditions are equivalent. }

Similarly, the only non trivial lines supported on the symmetry corner between $\mathcal{B}^{\text{sym}}=\mathcal{B}_1$ and $\mathcal{B}^{\text{bdy}}=\mathcal{B}_\infty$ are $L_{(\frac{\al}{2\pi},0)}$, with $\al\sim\al+2\pi$. The symmetry corner in this case is labeled by a parameter valued $S^1$: $\underline{B}_\al=[L_{(\frac{\al}{2\pi},0)}]$.\footnote{We abuse the notation a litter here. The corner $\underline{B}_\al$ should be understood in the context which $\mathcal{B}^{\text{bdy}}$ is given as in footnote \ref{footnote1}.} The fusion rules with symmetry lines are given by
\begin{align}
    L_{(\frac{\al'}{2\pi},0)} \otimes \underline{B}_\al=\underline{B}_{\al'+\al} \quad \text{and}\quad 
    L_{(0, \frac{\beta}{2\pi})}\otimes \underline{B}_\al=\underline{B}_{\al}, 
\end{align}
so the $\mathcal{B}^{\text{bdy}}=\mathcal{B}_\infty$ boundary realizes the $U(1)^m$-preserving-$U(1)^w$-breaking boundary states multiplet. We claim that this construction realizes the Neumann boundary states.

For the situation $\mathcal{B}^{\text{bdy}}=\mathcal{B}_{r>0}$, we first consider $r$ to be rational:
\begin{align}
    r=\frac{q}{p}\in\mathbb{Q}, \quad \gcd(p,q)=1.
\end{align}
Now, the holonomy of $\om$ along the symmetry corner takes value in $2\pi\frac{q}{p}\mathbb{Z}\cap 2\pi\mathbb{Z}=2\pi q\mathbb{Z}$, and the holonomy of $\widetilde{\om}$ along the corner takes value in $2\pi\frac{p}{q}\mathbb{Z}\cap 2\pi\mathbb{Z}=2\pi p\mathbb{Z}$. Thus, the line $L_{(\frac{n}{q},\frac{m}{p})}$, $n,m\in\mathbb{Z}$, is identified with the identity line. Note that the line $L_{(\frac{1}{q},\frac{1}{p})}=U^m_{\frac{2\pi}{p}}U^w_{\frac{2\pi}{p}}$ generates the $\mathbb{Z}_p^m\times\mathbb{Z}_q^w$ subgroup. So, the $\mathcal{B}^{\text{bdy}}=\mathcal{B}_{\frac{q}{p}}$ boundary realizes the $\mathbb{Z}_p^m\times\mathbb{Z}_q^w$-preserving boundary states multiplet.

However, we have shown that $\mathcal{B}_\frac{q}{p}$ do not satisfy the \textbf{conformal condition} when there exist non-zero integers $(n,m)$ satisfying $\frac{n}{rR}\pm mrR\notin \bZ$. For any $R\neq \frac{1}{r}=\frac{p}{q}$, one can find such $(n,m)$.\footnote{ 
When $rR=1$, $\frac{n}{rR}\pm m rR\in \bZ$, so there does not exist any desired $(n,m)$. When $rR= u\neq 1$ is an integer, we can take $(n,m)=(1,1)$ because $\frac{1}{rR} \pm rR=\frac{1}{u} \pm u \notin \bZ$. When $rR= \frac{u}{v}$ is a fractional number, we can take $(n,m)=(u,1)$ because $\frac{u}{rR}\pm rR= v\pm \frac{u}{v}\notin \bZ$. Then we are left with the case where $rR$ is an irrational number. We can first try $(n,m)=(1,1)$. If both $\frac{1}{rR}\pm rR\notin \bZ$, we are done. Otherwise, either (1) $\frac{1}{rR}+rR\in \bZ$ or (2) $\frac{1}{rR}-rR\in \bZ$. For case (1), we assume $\frac{1}{rR}+ rR=k \in \bZ$, and we can alternatively take $(n,m)=(1,2)$, so $\frac{1}{rR}+2rR= k+rR\notin\bZ$, and $\frac{1}{rR}-2rR= k- 3rR\notin \bZ$. For case (2), the proof is similar. 
}  Thus, we conclude that there do not exist simple conformal boundaries for compact boson CFT with radius $R\neq \frac{p}{q}$ to preserve $\mathbb{Z}_p^m\times\mathbb{Z}_q^w$ subgroup symmetry. In particular, any compact boson with irrational radius excludes such boundaries. Our result is consistent with the known boundary conditions and the symmetries they preserve as summarized in Table \ref{tab:R=u/v} and \ref{tab:Rirrational}.

\begin{table}\label{tab bdy states}
\centering
\begin{tabular}{|c|c|c|c|}
    \hline
    $\displaystyle \mathcal{B}^{\text{bdy}}$ candites& Conformal conditon & Symmetry preserved&Boundary state\\
    \hline
    $\displaystyle \mathcal{B}_0$& $\displaystyle\checkmark$ &$\displaystyle U(1)^w$ &$\displaystyle\ket{\beta}_{\text{Dir}}$\\
    \hline
    $\displaystyle \mathcal{B}_\infty$& $\displaystyle\checkmark$ & $\displaystyle U(1)^m$&$\displaystyle\ket{\al}_{\text{Neu}}$\\
    \hline
    $\displaystyle \mathcal{B}_{\frac{q}{p}}$& $\displaystyle\times$ iff $R\neq \frac{p}{q}$ & $\displaystyle\mathbb{Z}_p^m\times\mathbb{Z}_q^w$&$\displaystyle\times$ iff $R\neq \frac{p}{q}$\\
    \hline
\end{tabular}
\caption{Simple conformal boundary states of compact boson via SymTFT}
\end{table}

We added, that for irrational $r$ in $\mathcal{B}_r$, the holomomy of $\om$ along the corner takes value in $2\pi r \mathbb{Z}\cap 2\pi\mathbb{Z}=0$. It means that the topological junction $\underline{B}$ between $\mathcal{B}_r$ with irrational $r$ and $\mathcal{B}^{\text{sym}}$ does not preserve any non-trivial subgroup of $U(1)^m\times U(1)^w$. 

In conclusion, we have used our criteria to show that at a generic radius, the anomaly-free-$\mathbb{Z}_p^m\times\mathbb{Z}_q^w$ symmetry cannot be preserved by a simple conformal boundary state when $R\neq \frac{p}{q}$. In this case the lore is violated.

\section{The Lore Depends: Non-invertible Symmetries in Minimal Models}\label{sec: min model}

In this section, we discuss another example of CFT--the minimal model with non-invertible symmetry $\operatorname{Rep}(S_3)$--where the lore is violated. This example was first realized in \cite{Choi:2023xjw}. We also provide a CFT with $\operatorname{Rep}(S_3)$ symmetry--gauged coupled minimal models--where there exists symmetry conformal boundary conditions.

\subsection{The Lore Violated: Non-invertible Symmetries in Diagonal Minimal Models}\label{sec: min model M(6,5)}

In the unitary minimal model $M(m+1,m)$ with diagonal modular invariance, we make use of the following two facts:
\begin{enumerate}
    \item \textbf{Symmetry:} All simple symmetry lines satisfying operator/defect duality are Verlinde lines $\mathcal{L}_i$'s \cite{Verlinde:1988sn};
    \item \textbf{Boundary:} All simple conformal boundary states are Cardy states $\ket{B_i}$'s \cite{Cardy:1989ir,Cardy:2004hm}.
\end{enumerate}
Despite known in the literature, we revisit the second statement and provide a proof in Appendix \ref{cardy}.

It has been observed in \cite{Choi:2023xjw} that a non-invertible symmetry cannot be strongly preserved by any simple conformal boundary condition in minimal models. By definition, a strongly symmetric simple boundary state satisfies $\mathcal{L}_i \ket{B} = d_i \ket{B}$. While Cardy states satisfy  $\mathcal{L}_i \ket{B_j} = N_{ij}^k \ket{B_k}$ with $N_{ij}^k = 0$ or $1$, which is incompatible with some line $\mathcal{L}_i$ being non-invertible, i.e. $d_i > 1$.

This result does not depend on whether the symmetry is anomalous or not. As noted in \cite{Choi:2023xjw}, the $c=\frac{4}{5}$ tetracritical Ising model $M(6,5)$ has a $\operatorname{Rep}(S_3)$ sub-symmetry generated by $1 = \mathcal{L}_{\phi_{1,1}}$, $\eta = \mathcal{L}_{\phi_{1,5}}$, and $M = \mathcal{L}_{\phi_{1,3}}$, where $M$ is a non-invertible line with $d_M = 2$. The subcategory $\operatorname{Rep}(S_3)$, which is anomaly-free\footnote{The category $\operatorname{Rep}(G)$ admits a fiber functor to $\operatorname{Vec}$ by forgetting the group action on the vector space associated with each representation of $G$.}, is not preserved by any simple boundary condition, providing a counterexample to the strong version of the lore. However, the weak version of the lore is still obeyed.

In this subsection, we reinterpret the above results using the criteria proposed in Section \ref{sec: criteria}.

We begin with a minimal model $M(m+1,m)$ with diagonal modular invariance, defined on a closed 2-manifold. Considering its symmetry category of Verlinde lines $\mathcal{C}$, we construct the corresponding SymTFT whose lines are described by $\mathcal{Z}(\mathcal{C}) = \mathcal{C} \boxtimes \bar{\mathcal{C}}$. For convenience, we denote $i \coloneqq \mathcal{L}_i$ in $\mathcal{C}$. The symmetry boundary is chosen to be the canonical topological boundary of $\mathcal{Z}(\mathcal{C})$ associated with the diagonal Lagrangian algebra $\mathcal{A}^{\text{diag}}= \oplus_{i\in \mathcal{C}} i \boxtimes \bar{i}$, denoted by $\mathcal{B}^{\text{sym}}\simeq \mathcal{B}^{\text{diag}} \simeq \mathcal{C}$ on which Verlinde lines are supported.  The corresponding forgetful functor is
\begin{align}
    F_{\mathcal{C}}(i \boxtimes \bar{j}) = i \otimes \bar{j}.
\end{align}
We denote the physical boundary by $\widetilde{M}_m$, and summarize the setup as
\begin{align}
    M(m+1,m) \rightsquigarrow (\mathcal{B}^{\text{diag}} \simeq \mathcal{C}, \; \mathcal{C} \boxtimes \bar{\mathcal{C}}, \; \widetilde{M}_m).
\end{align}

Since the SymTFT is constructed from the maximal symmetry of $M(m+1,m)$, there is only one conformal family in each subspace
\begin{align}\label{conf family in M}
    \mathcal{H}_{\mu = i \boxtimes \bar{j}} = V_i \otimes \bar{V}_j.
\end{align}

Following the procedure outlined in Section \ref{sec: criteria}, we require the classification of topological boundaries of $\mathcal{C} \boxtimes \bar{\mathcal{C}}$. Fortunately, this problem is equivalent to classifying modular invariant partition functions of minimal models, which is already known. Suppose $\mathcal{C} \boxtimes \bar{\mathcal{C}}$ admits a topological boundary $\mathcal{B}$. We may choose $\mathcal{B}$ as the symmetry boundary. After shrinking the sandwich
\begin{align}\label{min sandwich}
    (\mathcal{B}, \mathcal{C} \boxtimes \bar{\mathcal{C}}, \widetilde{M}_m),
\end{align}
we obtain a theory with Hilbert space
\begin{align}\label{non-diag hilb}
    \mathcal{H} = \bigoplus_{F_{\mathcal{B}}(\mu) \succ 1} \mathcal{H}_\mu = \bigoplus_{F_{\mathcal{B}}(i \boxtimes \bar{j}) \succ 1} V_i \otimes \bar{V}_j.
\end{align}
This means that by considering all possible topological boundary conditions $\mathcal{B}$ of $\mathcal{C} \boxtimes \bar{\mathcal{C}}$, we recover all modular invariant partition functions. Since the modular invariant partition functions of minimal models follow the ADE classification \cite{francesco2012conformal,Goddard:1986ee}, all Lagrangian algebras corresponding to topological boundaries of $\mathcal{C} \boxtimes \bar{\mathcal{C}}$ are known. We note that diagonal and non-diagonal minimal models with the same central charge share the same physical boundary, differing only by the choice of symmetry boundary.

With this in mind, the step 2 in the criteria on \textbf{conformal condition} can now be checked explicitly. Different representations of the chiral Virasoro algebra in a minimal model have distinct conformal weights, i.e. $h_i \neq h_j$ for $i \neq j$. For $\mu$ to be condensable on $\mathcal{B}$, the sector $\mathcal{H}_\mu$ must contain a spin-0 conformal family, which occurs if and only if $\mu = i \boxtimes \bar{i}$. Therefore, only the canonical boundary $\mathcal{B}^{\text{diag}} \simeq \mathcal{C}$ satisfies the \textbf{conformal condition}. Notably, this result is independent of the choice of symmetry boundary.

In Section \ref{sec: criteria}, we remarked that the criteria generally provide only a necessary condition. However, by comparing the conclusion above with the second fact stated at the beginning of this section, we see that the conformal physical corner between $\mathcal{B}^{\text{diag}} \simeq \mathcal{C}$ and the physical boundary $\widetilde{M}_m$ exists and is unique.

Now let us turn to step 3. For diagonal $M(m+1,m)$, the symmetry boundary is also $\mathcal{B}^{\text{diag}} \simeq \mathcal{C}$, meaning that the symmetry corner is an object in $\mathcal{C}$. In other words, the conformal boundary conditions of a diagonal minimal model form the left regular module of $\mathcal{C}$, and the symmetry lines act on the boundary states via the fusion rules of $\mathcal{C}$, which are precisely the properties of the Cardy states. As a direct consequence, the absence of strongly $\operatorname{Rep}(S_3)$-symmetric conformal boundary in $M(6,5)$ immediately follows from the non-invertibility of $\operatorname{Rep}(S_3)$. Moreover, the existence of weakly $\operatorname{Rep}(S_3)$-symmetric conformal boundary follows from the fusion rule $M\otimes M \succ M, \eta\otimes M =M$. 

As a by-product, the boundary conditions of non-diagonal minimal models are also classified. We replace the symmetry boundary $\mathcal{B}^{\text{diag}} \simeq \mathcal{C}$ in \eqref{min sandwich} with $\mathcal{B}$. Then, the symmetry corner is an object in the $(\mathcal{B}, \mathcal{C})$-bimodule category. In other words, simple conformal boundary conditions of a non-diagonal minimal model with Hilbert space \eqref{non-diag hilb} correspond one-to-one with simple objects in the $(\mathcal{B}, \mathcal{C})$-bimodule category.

\subsection{The Lore Holds: $\operatorname{Rep}(S_3)$ in Ising${}^3/S_3$}\label{sec: gauged min model}

In this subsection, we construct an explicit example showing that the non-invertible symmetry $\operatorname{Rep}(S_3)$ can indeed be preserved by a simple conformal boundary condition, so the lore holds. This is a special case of the symmetric orbifold construction discussed in \cite{Choi:2023xjw}, but we decide to include here to contrast with the previous example where the lore is violated.

Consider three copies of the Ising CFT, denoted by $\text{Ising}^3 = \text{Ising} \otimes \text{Ising} \otimes \text{Ising}$, whose local operators take the form 
\begin{align}\label{ising3 op}
    \phi_1 \phi_2 \phi_3, \qquad \text{where}~\phi_i \in \{1, \epsilon, \sigma\}.
\end{align}
The modular-invariant torus partition function is
\begin{align}
    Z = Z_{\text{Ising}}^3 = \sum_{i,j,k = 1, \epsilon, \sigma} |\chi_i|^2 |\chi_j|^2 |\chi_k|^2.
\end{align}

The symmetric group $S_3$ acts on $\text{Ising}^3$ by permuting the three copies. Since the theory is non-chiral, the $S_3$ symmetry is anomaly-free according to the theorem in Appendix A of \cite{Albert:2022gcs}. We may therefore gauge the $S_3$ symmetry to obtain a new CFT, denoted by $\text{Ising}^3/S_3$. The gauged theory admits the dual non-invertible symmetry $\operatorname{Rep}(S_3)$. In what follows, we show that $\text{Ising}^3/S_3$ admits simple conformal boundary conditions that preserve this $\operatorname{Rep}(S_3)$ symmetry.

The SymTFT describing $S_3$ or $\operatorname{Rep}(S_3)$ symmetry is the $S_3$ gauge theory, denoted by $\mathfrak{D}(S_3)$. The Lagrangian algebras of $\mathfrak{D}(S_3)$ have been classified, see for instance \cite{Bhardwaj:2023idu,Cong:2016ayp}, and are given by\footnote{We follow the notations in \cite{Bhardwaj:2023idu}. The first factor in each component represents the conjugacy class of $S_3$, and the second factor represents the irreducible representation of the centralizer of the conjugate class of $S_3$. }
\begin{align}\label{A1A2A3A4}
\begin{split}
    \mathcal{A}_1 &= ([\operatorname{id}],1) \oplus ([\operatorname{id}],P) \oplus 2([\operatorname{id}],E), \\ \mathcal{A}_2 &= ([\operatorname{id}],1) \oplus ([\operatorname{id}],P) \oplus 2([a],1), \\
    \mathcal{A}_3 &= ([\operatorname{id}],1) \oplus ([\operatorname{id}],E) \oplus ([b],+), \\ \mathcal{A}_4 &= ([\operatorname{id}],1) \oplus ([a],1) \oplus ([b],+).
\end{split}
\end{align}
The boundaries where $\mathcal{A}_1$ or $\mathcal{A}_2$ condense support $S_3$ symmetry lines, while those where $\mathcal{A}_3$ or $\mathcal{A}_4$ condense support $\operatorname{Rep}(S_3)$ symmetry lines. The SymTFTs for $\text{Ising}^3$ and $\text{Ising}^3/S_3$ on a closed manifold are
\begin{align}
    \text{Ising}^3 \rightsquigarrow &(\mathcal{B}(\mathcal{A}_1) \simeq \operatorname{Vec}_{S_3},\; \mathfrak{D}(S_3),\; \widetilde{\text{Ising}^3}), \\
    \text{Ising}^3/S_3 \rightsquigarrow &(\mathcal{B}(\mathcal{A}_4) \simeq \operatorname{Rep}(S_3),\; \mathfrak{D}(S_3),\; \widetilde{\text{Ising}^3}).
\end{align}

To find a boundary state in $\text{Ising}^3/S_3$ preserving the $\operatorname{Rep}(S_3)$ symmetry, we take $\mathcal{B}^{\text{bdy}} = \mathcal{B}(\mathcal{A}_1)$, since $\mathcal{B}(\mathcal{A}_1)$ is obtained from  $\mathcal{B}(\mathcal{A}_4)$ by gauging $\operatorname{Rep}(S_3)$. We now check the \textbf{conformal condition} on $\mathcal{B}(\mathcal{A}_1)$. For each $\mu \in \mathfrak{D}(S_3)$ condensable on $\mathcal{B}(\mathcal{A}_1)$ (that is, $\mu \prec \mathcal{A}_1$), the conformal families in the sector $\mathcal{H}_\mu$ are precisely those of $\text{Ising}^3$, as in \eqref{ising3 op}, all of which have spin~0. Hence, the criterion does not forbid the existence of such a state.

The criterion is merely a necessary condition. To reach a definitive conclusion, we must check whether $\mathcal{B}(\mathcal{A}_1)$ admits a conformal corner with $\widetilde{\text{Ising}^3}$. This can be analyzed in the ungauged theory $\text{Ising}^3$.

The simple conformal boundary states of $\text{Ising}^3$ that commute with three copies of Virasoro algebras are known: they are tensor products of the three Cardy states of the Ising CFT. We examine their behavior under the $S_3$ action and organize them into multiplets, listed in the first column of Table \ref{ising3 states}.

\begin{table}
\centering
\begin{tblr}{
  colspec = {X[c] X[c]},
  width = {10cm},
  vlines,
  hline{1,2,5,11,12} = {-}{},
  hline{3,4,6,7,8,9,10} = {1-1}{},
  row{1} = {font=\bfseries},
  rowsep = 2pt,
}
\textbf{Multiplets} & $\mathcal{B}^{\text{bdy}}$ \textbf{boundary} \\ 
$\ket{111}$ & \\
$\ket{\epsilon\epsilon\epsilon}$ & $\mathcal{B}(\mathcal{A}_4)$ \\
$\ket{\sigma\sigma\sigma}$ & \\
$\ket{11\epsilon},\ket{\epsilon11},\ket{1\epsilon1}$ & \\
$\ket{11\sigma},\ket{\sigma11},\ket{1\sigma1}$ & \\
$\ket{\epsilon\epsilon1},\ket{1\epsilon\epsilon},\ket{\epsilon1\epsilon}$ & $\mathcal{B}(\mathcal{A}_3)$ \\
$\ket{\epsilon\epsilon\sigma},\ket{\sigma\epsilon\epsilon},\ket{\epsilon\sigma\epsilon}$ & \\
$\ket{\sigma\sigma1},\ket{1\sigma\sigma},\ket{\sigma1\sigma}$ & \\
$\ket{\sigma\sigma\epsilon},\ket{\epsilon\sigma\sigma},\ket{\sigma\epsilon\sigma}$ &  \\
$\begin{array}{c}
\ket{1\epsilon\sigma},\ket{\epsilon1\sigma},\ket{1\sigma\epsilon},\\
\ket{\sigma\epsilon1},\ket{\sigma1\epsilon},\ket{\epsilon\sigma1}
\end{array}$ & $\mathcal{B}(\mathcal{A}_1)$  \\
\end{tblr}
\caption{Multiplets of boundary states in the $\text{Ising}^3$ theory and their associated $\mathcal{B}^{\text{bdy}}$ boundaries.}
\label{ising3 states}
\end{table}

To associate each multiplet with its corresponding $\mathcal{B}^{\text{bdy}}$ boundary, we analyze the symmetry corners between $\mathcal{B}^{\text{sym}} = \mathcal{B}(\mathcal{A}_1)$ and the other topological boundaries $\mathcal{B}(\mathcal{A}_i)$. 

For interfaces between distinct topological boundaries, we use the following fact: the number of simple objects in the $(\mathcal{B}(\mathcal{A}_i), \mathcal{B}(\mathcal{A}_j))$-bimodule category equals the dimension of the space $\operatorname{Hom}_{\mathcal{Z}(\mathcal{C})}(\mathcal{A}_i, \mathcal{A}_j)$ \cite{Putrov:2024uor,Bhardwaj:2023idu,Thorngren:2019iar}. From the Lagrangian algebras \eqref{A1A2A3A4}, we find
\begin{align}
\begin{split}
    &\dim_{\mathbb{C}} \operatorname{Hom}(\mathcal{A}_1, \mathcal{A}_1) = 6, \\
    &\dim_{\mathbb{C}} \operatorname{Hom}(\mathcal{A}_1, \mathcal{A}_2) = 2, \\
    &\dim_{\mathbb{C}} \operatorname{Hom}(\mathcal{A}_1, \mathcal{A}_3) = 3, \\
    &\dim_{\mathbb{C}} \operatorname{Hom}(\mathcal{A}_1, \mathcal{A}_4) = 1.
\end{split}
\end{align}
Since the number of simple boundary states equals the number of simple objects in the corresponding bimodule category, we obtain the assignments shown in the second column of Table \ref{ising3 states}. In particular, the last row shows that $\mathcal{B}(\mathcal{A}_1)$ indeed admits a conformal corner with $\widetilde{\text{Ising}^3}$.

We thus conclude that $\text{Ising}^3/S_3$ admits simple conformal boundary conditions that strongly preserve the $\operatorname{Rep}(S_3)$ symmetry.

\section{The Lore Unsettled: Non-invertible symmetries in WZW Models}\label{sec: su3 wzw}

In this section, we present a new example of a CFT with a strongly anomaly-free fusion category symmetry $\operatorname{Rep}(A_4)$. Even though there are no known boundary states strongly preserving this symmetry, our criteria do not forbid the existence of such symmetric simple conformal boundary state.

We emphasize that the boundary state we proposed is beyond the Cardy states, and the conclusions reached here do not rely on the complete classification of conformal boundary states of RCFT, which have not been done as far as we know. However, the SymTFT technique completely fixes the symmetry action on conformal boundary states.

Finally, we make a remark on Tetracritical Ising category symmetry in Section \ref{sec: remark M(6,5)}, which can be realized as a subcategory of the Verlinde lines category in the $SU(2)_1\times SU(2)_3\times SU(2)_{-4}$ WZW model. In this theory, our criteria suggests a non-Cardy, $\operatorname{Rep}(S_3)$ symmetry preserving simple conformal boundary state.

\subsection{$\operatorname{Rep}(A_4)$ symmetry in $SU(3)_3$ WZW model}
Consider the $c=4$ $SU(3)_3$ WZW model with diagonal modular invariance. We denote the category of its Verlinde lines as $\mathcal{C}$. The Verlinde lines may be labeled by
\begin{align}
\begin{array}{cccccccccc}\label{verlinde in su(3)3}
    (0,0)_0&(3,0)_1&(0,3)_1&(1,1)_{\frac{1}{2}}&(1,0)_{\frac{2}{9}}&(2,1)_{\frac{8}{9}}&(0,2)_{\frac{5}{9}}&(0,1)_{\frac{2}{9}}&(2,0)_{\frac{5}{9}}&(1,2)_{\frac{8}{9}} \\
     1&\tau&\tau^2&X&Y&\tau Y&\tau^2Y&Z&\tau Z&\tau^2Z.
\end{array}
\end{align}
where the first row is the standard Dynkin label, with the subscripts being conformal weights $h$'s of the corresponding chiral algebra primaries. The fusion rules are
\begin{align}
\tau^3 &= 1, 
& \tau X &= X, 
& X^2 &= 1 \oplus \tau \oplus \tau^2 \oplus 2X, \notag\\[4pt]
Y^2 &= Z \oplus \tau Z, 
& Z^2 &= Y \oplus \tau^2 Y, \notag\\[4pt]
XY &= Y \oplus \tau Y \oplus \tau^2 Y, 
& XZ &= Z \oplus \tau Z \oplus \tau^2 Z, 
& YZ &= 1 \oplus X.
\end{align}
Here $\tau$ is an invertible line of order 3 generating the center $\mathbb{Z}_3$ symmetry. The lines $X$, $Y$ and $Z$ are non-invertible topological lines with quantum dimension $d_X=3$ and $d_Y=d_Z=2$ respectively. \footnote{Note that in this theory all Verline lines has integer quantum dimensions.}

We note that $1,\tau,\tau^2, X$ are closed under fusion, forming a fusion subcategory, denoted as $\mathcal{D}$. Furthermore, since Verlinde lines of an RCFT form a modular tensor category, the subcategory $\mathcal{D}$ is braided. It turn out that its braiding data (i.e. $S$ and $T$ matrices) together with fusion rules significantly constrain its property. We claim that
\begin{align}\label{D rep}
    \mathcal{D}\xrightarrow[\text{forget braiding}]{\simeq} \operatorname{Rep}(A_4).
\end{align}
We leave the proof in Appendix \ref{near group}, together with some relevant modular data.

Since the $\operatorname{Rep}(A_4)$ category has a fiber functor to $\operatorname{Vec}$, the subsymmetry $\mathcal{D}$ is strongly anomaly-free.

\subsection{Criteria Applied}
In this subsection, we show that a particular topological boundary of SymTFT, chosen as $\mathcal{B}^{\text{bdy}}$, satisfies both the \textbf{conformal condition} and strongly $\operatorname{Rep}(A_4)$-\textbf{symmetric condition}.

We construct the SymTFT $\mathcal{Z}(\mathcal{C})=\mathcal{C}\boxtimes \bar{\mathcal{C}}$ from the Verlinde lines symmetry $\mathcal{C}$. Note that the modular invariant partition function of $SU(3)_3$ WZW model, written in terms of $su(3)_3$ chiral algebra characters $\chi_\lam(q)$'s, are equivalent to the Lagrangian algebra of $\mathcal{Z}(\mathcal{C})$ \cite{francesco2012conformal}, 
\begin{align}
    Z=\sum_{\lam,\mu}M_{\lam\mu}\chi_{\lam}(q)\bar{\chi}_{\mu}(\bar{q}) \quad \leftrightsquigarrow \quad \mathcal{A}=\sum_{\lam,\mu}M_{\lam\mu} (\lam\boxtimes\bar{\mu}),
\end{align}
and correspond to topological boundary conditions of $\mathcal{Z}(\mathcal{C})$.

The partition function of $SU(3)_3$ has been classified in \cite{Gannon:1992ty, Gannon:1993nk}, which are 
\begin{align}
    &Z_1=\sum_{\lam} |\chi_\lam|^2\label{su3 z1},\\
    &Z_2=\sum_{\lam=(\lam_1,\lam_2)}\chi_{(\lam_1,\lam_2)}\bar{\chi}_{(\lam_2,\lam_1)}\label{su3 z2},\\
    &Z_3=|\chi_1+\chi_\tau+\chi_{\tau^2}|^2+3|\chi_X|^2\label{su3 z3}.
\end{align}
Here we abuse the notation by identifying the Dykin label with the label of Verline lines. We denote the corresponding Lagrangian algebras as $\mathcal{A}_1,\mathcal{A}_2,\mathcal{A}_3$, and corresponding topological boundaries as $\mathcal{B}_1,\mathcal{B}_2,\mathcal{B}_3$ respectively. Note that in the model we considered, which is diagonal WZW, we have $\mathcal{B}^{\text{sym}}\simeq\mathcal{B}_1\simeq\mathcal{C}$.

We check the \textbf{conformal condition} first. In a representation $V_\lam\otimes\bar{V}_\mu$ of $su(3)_k$ chiral algebra, Virasoro primaries are obtained by applying $J^a_n$'s and $\bar{J}^a_n$'s on the $su(3)_k$ primary state $\ket{\lam,\mu}$. By the commutation relation $[L_n,J^a_m]=-mJ^a_{n+m}$, we know that the conformal weights of Virasoro primaries in $V_\lam$ take values in 
\begin{align}\label{conf weight in WZW}
    (h,\bar{h})\in(h_\lam+\mathbb{Z}_{\geq0}, h_\mu+\mathbb{Z}_{\geq0}).
\end{align}
All the boundaries $\mathcal{B}_1, \mathcal{B}_2$ and $\mathcal{B}_3$ satisfy the conformal condition according to \eqref{verlinde in su(3)3}, especially the property $h_{(\lam_1,\lam_2)}=h_{(\lam_2,\lam_1)}$.

We claim that the topological boundary $\mathcal{B}_3$ corresponds to gauging $\operatorname{Rep}(A_4)$ from $\mathcal{B}_1$, hence the $\operatorname{Rep}(A_4)$-\textbf{symmetric condition} is satisfied by $\mathcal{B}_3$ following Section \ref{sec: criteriasub}.

To prove the claim we need to show that the partition function $Z_3$ \eqref{su3 z3} is obtained by gauging $\operatorname{Rep}(A_4)$ from the diagonal $SU(3)_3$ WZW $Z_1$ \eqref{su3 z1}. The Frobenius algebra corresponds to $\operatorname{Rep}(A_4)$ is $A(\operatorname{Rep}(A_4))=1\oplus\tau\oplus\tau^2\oplus3X$. However, the algebra $A(\operatorname{Rep}(A_4))$ is Morita equivalent to the Frobenius algebra of $\mathbb{Z}_3$: $A(\mathbb{Z}_3)=1\oplus\tau\oplus\tau^2$. This follows from eq.(3.27) in \cite{Diatlyk:2023fwf} and
\begin{align}
    A(\operatorname{Rep}(A_4))=Y\otimes A(\mathbb{Z}_3)\otimes Z, \quad \text{where}\, Z=\bar{Y}.
\end{align}
Since $\mathbb{Z}_3$ is cyclic, by using the modular data \eqref{S WZW} and 
\begin{align}
    &Z_{(g,1)}(\tau)=\sum_{\lam}\frac{S_{g\lam}}{S_{1\lam}}\chi_{\lam}(-\frac{1}{\tau})\bar{\chi}_{\lam}(-\frac{1}{\bar{\tau}}),\\
    &Z_{(g,h)}(\tau)=Z_{(gh,h)}(\tau+1),
\end{align}
where $\tau$ is the modular parameter, and $Z_{(g,h)}$ is a torus partition function with $g$-twist along the time cycle and $h$-twist along the space cycle, we can show that $Z_3$ can be obtained by gauging $\mathbb{Z}_3$ from $Z_1$:
\begin{align}
    Z_3(\tau)=\frac{1}{3}\sum_{g,h\in\mathbb{Z}_3}Z_{(g,h)}(\tau).
\end{align}
Morita equivalent Frobenius algebras gives physically equivalent gauging, thus we have proved the claim.

Choosing $\mathcal{B}^{\text{bdy}}=\mathcal{B}_3$ and the symmetry corner $\underline{B}=A(\operatorname{Rep}(A_4))\in\mathcal{C}_{A(\operatorname{Rep}(A_4))}$ following Section \ref{sec: criteriasub}, the $\operatorname{Rep}(A_4)$ symmetry is strongly preserved. Since we cannot show that the criteria are sufficient, we still cannot firmly conclude that such conformal boundary condition do exist. In this case the lore is still unsettled.

Note that the Cardy states of $SU(3)_3$ WZW correspond to choosing $\mathcal{B}^{\text{bdy}}=\mathcal{B}_1$, and none of them strongly preserves $\operatorname{Rep}(A_4)$. Of course, the Cardy state $\ket{X}$ is weakly symmetric under $\operatorname{Rep}(A_4)$.

\subsection{Revisiting the Tetracritical Ising Symmetry}\label{sec: remark M(6,5)}

In this subsection, we revisit the tetracritical Ising category symmetry, which is the modular tensor category formed by Verline lines in the diagonal unitary minimal model $M(6,5)$. As discussed in Section \ref{sec: min model M(6,5)}, we showed that, in the $M(6,5)$ minimal model, there is no simple $\operatorname{Rep}(S_3)$-preserving conformal boundary condition. If the \textbf{Improved Lore} were to hold, there should be another CFT with the tetracritical Ising category symmetry where a simple $\operatorname{Rep}(S_3)$-preserving conformal boundary condition exists. In this subsection, we propose that a candidate of such CFT is the diagonal $SU(2)_1\times SU(2)_3\times SU(2)_{-4}$ WZW model.\footnote{This model is motivated by the coset construction of the minimal model $M(6,5)= (SU(2)_1\times SU(2)_3)/SU(2)_4$.}

We denote the symmetry category of Verlinde lines in $SU(2)_k$ WZW as $(A_1, k)$ following \cite{Choi:2023xjw}, with $\mathcal{L}_j\in(A_1, k)$ for $SU(2)$ spin $j=0,\frac{1}{2},\cdots,\frac{k}{2}$.

Consider the diagonal $SU(2)_1\times SU(2)_3\times SU(2)_{-4}$ WZW model, with symmetry $(A_1,1)\boxtimes(A_1,3)\boxtimes\overline{(A_1,4)}$. We claim the following subcategory relation
\begin{align}\label{subcategories}
    \operatorname{Rep}(S_3)\subset \mathsf{tetracritical}\hspace{0.5em}\mathsf{Ising}\subset(A_1,1)\boxtimes(A_1,3)\boxtimes\overline{(A_1,4)}
\end{align}
with the identification
\begin{align}
\begin{array}{c|c|c|c|c}
    1&\eta&M&W&N\\
    \mathcal{L}_0\boxtimes\mathcal{L}_0\boxtimes\overline{\mathcal{L}_0}
    &\mathcal{L}_0\boxtimes\mathcal{L}_0\boxtimes\overline{\mathcal{L}_2}
    &\mathcal{L}_0\boxtimes\mathcal{L}_0\boxtimes\overline{\mathcal{L}_1}
    &\mathcal{L}_0\boxtimes\mathcal{L}_1\boxtimes\overline{\mathcal{L}_0}
    &\mathcal{L}_{\frac{1}{2}}\boxtimes\mathcal{L}_0\boxtimes\overline{\mathcal{L}_{\frac{3}{2}}}\\
    \hline
    MW&\eta W&\eta N&WN&\eta WN\\
    \mathcal{L}_0\boxtimes\mathcal{L}_1\boxtimes\overline{\mathcal{L}_1}
    &\mathcal{L}_{\frac{1}{2}}\boxtimes\mathcal{L}_{\frac{1}{2}}\boxtimes\overline{\mathcal{L}_1}
    &\mathcal{L}_0\boxtimes\mathcal{L}_{\frac{3}{2}}\boxtimes\overline{\mathcal{L}_{\frac{3}{2}}}
    &\mathcal{L}_{\frac{1}{2}}\boxtimes\mathcal{L}_1\boxtimes\overline{\mathcal{L}_{\frac{3}{2}}}
    &\mathcal{L}_0\boxtimes\mathcal{L}_{\frac{1}{2}}\boxtimes\overline{\mathcal{L}_{\frac{3}{2}}}
\end{array}
\end{align}
where we label lines in $\mathsf{tetracritical}\hspace{0.5em}\mathsf{Ising}$ following \cite{Chang:2018iay}. The relation \ref{subcategories} can be check by calculate the S and T sub-matrices of 10 objects listed above. The lines $\{1,\eta, M\}$ generate $\operatorname{Rep}(S_3)$.

Now consider the boundary SymTFT for the diagonal $SU(2)_1\times SU(2)_3\times SU(2)_{-4}$ WZW model whose lines form the Drinfeld center of $(A_1,1)\boxtimes(A_1,3)\boxtimes\overline{(A_1,4)}$. The symmetry boundary $\mathcal{B}^{\text{sym}}$ corresponds to the diagonal Lagrangian algebra of the bulk TQFT. Consider the topological boundary obtained by gauging the anomaly-free $\operatorname{Rep}(S_3)$ from the symmetry boundary: $\mathcal{B}=\mathcal{B}^{\text{sym}}/\operatorname{Rep}(S_3)$. For the same reason discussed around \ref{conf weight in WZW}, the boundary $\mathcal{B}$ satisfies the conformal condition. Thus our criteria do not forbid the existence of $\operatorname{Rep}(S_3)$ strongly symmetric simple conformal boundary state in this model, similar to the situation in the previous case. Moreover, if such boundary condition exist, it must be one that does not commute with chiral algebra, which would be interesting to construct in the future.

\section{Conclusion and Future Directions}

In this work, we proposed criteria for the existence of conformal boundary conditions with prescribed symmetry properties using the symmetry TFT framework. We demonstrated that even for an anomaly-free global symmetry, there can exist obstructions preventing its realization by a simple conformal boundary condition in 1+1d CFTs. A practical procedure was introduced to detect such obstructions by incorporating the minimal dynamical data, namely the conformal weights of Virasoro primary operators. Four detailed examples were provided to illustrate the procedure.

Below we mention several future directions and open questions.
\begin{enumerate}
	\item One open question concerns the validity of the lore of the weak version. To the authors' best knowledge, when $\mathcal{D}$ is weakly anomaly free but not strongly anomaly free, we are not aware of a CFT where a simple weakly-symmetric conformal boundary condition is absent. It would be interesting to either find such an example or prove that such case is always forbidden.
	\item Since the lore in Section \ref{sec: su3 wzw} is unsettled, and we haven't found obstructions for a simple $\operatorname{Rep}(A_4)$-symmetric conformal boundary condition in $SU(3)_3$ WZW model as well as a simple $\operatorname{Rep}(S_3)$-symmetric conformal boundary condition in $SU(2)_1\times SU(2)_3 \times SU(2)_{-4}$ WZW model, it is very likely that such boundary conditions exist, and it would be desirable to construct such boundary conditions explicitly.
	\item Since this work discusses only boundary conditions in 2d CFTs, it would be interesting to investigate the lore in higher dimensional CFTs.
    \item We showed that in certain theories, conformal condition and symmetric condition can not be both satisfied for boundary conditions. This suggests certain mixed anomaly between the conformal symmetry and internal (non-invertible) symmetry in the CFTs. It would be nice to formulate such mixed anomaly more explicitly.\footnote{We thank Po-Shen Hsin for pointing out this observation. }
\end{enumerate}

\section*{Acknowledgments}

We are grateful to Philip Boyle Smith, Yichul Choi, Po-Shen Hsin, Justin Kaidi, Yuya Kusuki, Zohar Komargodski, Yuya Kusuki,  Ho Tat Lam, Kantaro Ohmori, Brandon Rayhaun, Shu-Heng Shao, Yi-Nan Wang for helpful discussions, and to Yichul Choi, Brandon Rayhaun, Zohar Komargodski and Shu-Heng Shao for helpful comments on a draft. We also thank the Kavli Institute for Theoretical Physics (KITP) for hospitality during the program GenSym25, during which this project was initiated. This research was supported in part by grant NSF PHY-2309135 to the KITP. The work of Y.Z. is supported by NSFC grant No.12505093 and the starting funds from University of Chinese Academy of Sciences (UCAS) and from the Kavli Institute for Theoretical Sciences (KITS).

\appendix

\section{The Lore Holds: Review of Examples}
\label{app:review}

There are many examples in which the lore is obeyed. 

One of the most well known family of examples is 2d $N$ Dirac fermions with anomalous $U(1)^{2N}$ symmetries. It is known in \cite{Smith:2019jnh,Smith:2020nuf,Smith:2020rru,Wang:2013yta,Wang:2022ucy,Zeng:2022grc,Wang:2018ugf,vanBeest:2023dbu} that the $U(1)^N$ subgroup is anomaly free, once the charges of fermions are carefully chosen, among which the 3-4-5-0 model is the most well known. The simple $U(1)^N$ symmetric conformal boundary states have been explicitly constructed in \cite{Smith:2019jnh,Smith:2020nuf,Smith:2020rru}. 

Another class of examples with finite invertible symmetries is the diagonal WZW models. In \cite{Li:2022drc,Numasawa:2017crf,Gaberdiel:2002qa,Aharony:2016jvv,Kikuchi:2019ytf}, simple Cardy states have been found to preserve any anomaly free subgroup that commutes with the chiral algebra, hence support the lore. Below we review the anomalies of the invertible symmetries in WZW theories, and check that the anomaly-free condition coincides with the condition for symmetric Cardy states. 

\subsection*{$SU(N)_k$ WZW}

The invertible symmetry commuting with the $SU(N)\times SU(N)$ chiral algebra is $\bZ_N$. The anomaly is $k$ mod $N$. The anomaly-free subgroup is $\bZ_L$ if $N/L \in \bZ$ and $k/L \in 
\bZ$, hence $L= 0$ mod $\gcd(k,N)$.

The Cardy states are labeled by representations of the $su(N)_k$ affine Lie algebra. Denote the Cardy state as
\begin{equation}\label{eq:DynkinConstraint}
    \ket{[\lambda_0; \lambda_1, ..., \lambda_{N-1}]}, \quad \text{with} \quad k= \sum_{i=0}^{N-1} \lambda_i.
\end{equation}
The $\bZ_N$ generator acts on the representation as
\begin{equation}
    \bZ_N: [\lambda_0; \lambda_1, ..., \lambda_{N-1}]\to  [\lambda_{N-1}; \lambda_0, \lambda_1, ..., \lambda_{N-2}].
\end{equation}
The generator of the anomaly free $\bZ_L= \bZ_{\gcd(k,N)}$ subgroup acts on the representation as 
\begin{equation}
    \bZ_{\gcd(k,N)}: [\lambda_0; \lambda_1, ..., \lambda_{N-1}] \to [\lambda_{N- \frac{N}{\gcd(k,N)}}; \lambda_{N- \frac{N}{\gcd(k,N)}+1}, ..., \lambda_{N- \frac{N}{\gcd(k,N)}-1}].
\end{equation}
Requiring the Cardy state to be $\bZ_{\gcd(k,N)}$ symmetric means 
\begin{equation}
    \ket{[\lambda_0; \lambda_1, ..., \lambda_{N-1}]} =\ket{[\lambda_{N- \frac{N}{\gcd(k,N)}}; \lambda_{N- \frac{N}{\gcd(k,N)}+1}, ..., \lambda_{N- \frac{N}{\gcd(k,N)}-1}]} .
\end{equation}
This imposes $\lambda_i = \lambda_{i- \frac{N}{\gcd(N,k)}}$, where the subscript is defined mod $N$. Feeding this constraint into \eqref{eq:DynkinConstraint}, we obtain 
\begin{equation}
    \sum_{i=0}^{\frac{N}{\gcd(N,k)}-1} \lambda_i = \frac{k}{\gcd(k,N)},
\end{equation}
where both sides are integers. It is obvious that there is no Cardy state preserving $\bZ_L$ with $L> \gcd(N,k)$, since it would render the RHS above to be a non-integer. In summary, we indeed find a symmetric Cardy state preserving any anomaly free subgroup of $\bZ_N$.

\subsection*{$Spin(2N+1)_k$ WZW}

The center symmetry is $\bZ_2$, and is always anomaly free for any $k$. 

The Cardy states are labeled by the representation of the $so(N)_k$ affine Lie algebra. Denote the Cardy state as 
\begin{equation}\label{eq:dynkinspin2np1}
    \ket{[\lambda_0; \lambda_1, ..., \lambda_N]}, \quad \text{with} \quad k= \lambda_0 + \lambda_1 + \lambda_N+ 2\sum_{i=2}^{N-1}\lambda_i.
\end{equation}
The $\bZ_2$ generator acts on the representation as 
\begin{equation}
    \bZ_2: [\lambda_0; \lambda_1, ..., \lambda_N] \to [\lambda_1; \lambda_0, \lambda_2, ..., \lambda_N].
\end{equation}
Requiring the Cardy state transforms trivially under the anomaly free $\bZ_2$, we get $\lambda_0=\lambda_1$. Substituting it into the level condition \eqref{eq:dynkinspin2np1}, we get 
\begin{equation}
    k= \lambda_N + 2\sum_{i=1}^{N-1} \lambda_i,
\end{equation}
which can always be satisfied. In summary, we indeed find a symmetric Cardy state preserving any anomaly free symmetry $\bZ_2$.

\subsection*{$Spin(4N+2)_k$ WZW}

The center symmetry is $\bZ_4$. The anomaly is $k$ mod $4$. Hence the anomaly free subgroup is $\bZ_4$ for $k\in 4\bZ$, $\bZ_2$ for $k\in 4\bZ+2$, and $\bZ_1$ for $k\in 2\bZ+1$. 

The Cardy states are 
\begin{equation}
    \ket{[\lambda_0; \lambda_1, ..., \lambda_{2N+1}]}\quad \text{with} \quad k= \lambda_0 + \lambda_1 + \lambda_{2N} + \lambda_{2N+1} + 2 \sum_{i=2}^{2N-1} \lambda_i.
\end{equation}
The $\bZ_4$ generator acts on the representation as 
\begin{equation}
    \bZ_4: [\lambda_0; \lambda_1, ..., \lambda_{2N+1}] \to [\lambda_{2N}; \lambda_{2N+1}, \lambda_{2N-1}, \lambda_{2N-2}, ..., \lambda_1, \lambda_0].
\end{equation}
Requiring the Cardy state to be $\bZ_4$ invariant, we have $\lambda_0=\lambda_1=\lambda_{2N} = \lambda_{2N+1}$ and $\lambda_{i} = \lambda_{2N+1-i}$. Substituting into the level condition we find 
\begin{equation}
    k= 4\sum_{i=0}^{N} \lambda_i,
\end{equation}
which requires $k\in 4\bZ$. 

If we only demand $\bZ_2\subset \bZ_4$ invariance generated by 
\begin{equation}
    \bZ_2: [\lambda_0; \lambda_1, ..., \lambda_{2N+1}] \to [\lambda_1; \lambda_0, \lambda_2, ..., \lambda_{2N-1}, \lambda_{2N+1}, \lambda_{2N}],
\end{equation}
we have $\lambda_1=\lambda_0, \lambda_{2N}= \lambda_{2N+1}$. Substituting into the level condition we find 
\begin{equation}
    k= 2(\lambda_0 + \lambda_{2N}) + 2 \sum_{i=2}^{2N-1}\lambda_i,
\end{equation}
which requires $k\in 2\bZ$. 

Finally when $k\in 2\bZ+1$, there is no Cardy state preserving any non-trivial subgroup of $\bZ_4$. These conditions precisely match the anomaly-free condition.

\subsection*{$Sp(N)_k$ WZW}

The center symmetry is $\bZ_2$, and is anomaly free when $kN\in 2\bZ$. 

The Cardy states are labeled by the representation of the $sp(N)_k$ affine Lie algebra. Denote the Cardy state as 
\begin{equation}\label{eq:dynkinspn}
    \ket{[\lambda_0; \lambda_1, ..., \lambda_N]}, \quad \text{with} \quad k= \sum_{i=0}^{N}\lambda_i.
\end{equation}
The $\bZ_2$ generator acts on the representation as 
\begin{equation}
    \bZ_2: [\lambda; \lambda_1, ..., \lambda_N] \to [\lambda_N; \lambda_{N-1}, ..., \lambda_0].
\end{equation}
Requiring the Cardy state to be invariant under $\bZ_2$, we get $\lambda_i=\lambda_{N-i}$. Substituting it into the level condition \eqref{eq:dynkinspn}, we get 
\begin{equation}
    k= 
    \begin{cases}
        \lambda_{\frac{N}{2}} + 2\sum_{i=0}^{\frac{N}{2}-1} \lambda_i, & N\in 2\bZ\\
        2\sum_{i=0}^{\frac{N-1}{2}} \lambda_i, & N\in 2\bZ+1\\
    \end{cases}
\end{equation}
which can always be satisfied by any $k$ for even $N$, and only by $k \in 2\bZ$ for odd $N$. Equivalently, level condition can be satisfied when $kN\in 2\bZ$, which precisely matches the anomaly free condition. In summary, we indeed find a symmetric Cardy state preserving any anomaly free subgroup of $\bZ_2$.

\subsection*{$(E_6)_k$ WZW}

The center symmetry is $\bZ_3$, and its anomaly is $k$ mod $3$. 

The Cardy state is
\begin{equation}
    \ket{[\lambda_0; \lambda_1, ..., \lambda_6]}\quad \text{with} \quad k= \lambda_0+ \lambda_1+ 2\lambda_2 + 3 \lambda_3 + 2 \lambda_4 + \lambda_5 + 2\lambda_6.
\end{equation}
The $\bZ_3$ acts on the representation as 
\begin{equation}
    \bZ_3: [\lambda_0; \lambda_1, ..., \lambda_6] \to [\lambda_1; \lambda_5, \lambda_4, \lambda_3, \lambda_6, \lambda_0, \lambda_2].
\end{equation}
Invariance under $\bZ_3$ implies $\lambda_0= \lambda_1 = \lambda_5$ and $\lambda_2= \lambda_4 = \lambda_6$. Substituting these into the level condition, we find 
\begin{equation}
    k= 3\lambda_0 + 3\lambda_3 + 6 \lambda_2,
\end{equation}
which requires $k=0$ mod 3. This is precisely the anomaly vanishing condition.

\subsection*{$(E_7)_k$ WZW}

The center symmetry is $\bZ_2$, and its anomaly is $k$ mod 2. 

The Cardy state is
\begin{equation}
    \ket{[\lambda_0; \lambda_1, ..., \lambda_7]}\quad \text{with} \quad k= \lambda_0+ 2\lambda_1+ 3\lambda_2 + 4 \lambda_3 + 3 \lambda_4 + 2\lambda_5 + \lambda_6+ 2\lambda_7.
\end{equation}
The $\bZ_3$ acts on the representation as 
\begin{equation}
    \bZ_2: [\lambda_0; \lambda_1, ..., \lambda_7] \to [\lambda_6; \lambda_5, \lambda_4, \lambda_3, \lambda_2, \lambda_1, \lambda_0, \lambda_7].
\end{equation}
Invariance under $\bZ_2$ implies $\lambda_0= \lambda_6$, $\lambda_1= \lambda_5$, and $\lambda_2= \lambda_4$. Substituting these into the level condition, we find 
\begin{equation}
    k= 2\lambda_0 + 4\lambda_1 + 6 \lambda_2 + 4 \lambda_3 + 2 \lambda_7,
\end{equation}
which requires $k=0$ mod 2. This is precisely the anomaly vanishing condition.

\section{Ishibashi's Results}\label{ishibashi}
In this section, we review Ishibashi’s proof \cite{Ishibashi:1988kg,Onogi:1988qk}\footnote{In Ishibashi's original paper \cite{Ishibashi:1988kg}, the proof is presented for the $SU(2)_k$ WZW model. However, the result can be generalized to (rational) CFTs as well.} of the following statement:  
in the closed-string channel, all solutions of the boundary conformal condition 
\begin{align}\label{bdy conformal}
    L_n \ishi{I} = \bar{L}_{-n} \ishi{I}
\end{align}
are spanned by the Ishibashi states.

The $a$-twisted Hilbert space takes the form\footnote{We do not specify the symmetry $\mathcal{C}\ni a$ here, because we only require that the Hilbert space adimit such direct sum decomposition.}
\begin{align}
    \mathcal{H}^a = \bigoplus_{ij} M^a_{ij} \, V_i \otimes \bar{V}_j,
\end{align}
where $M_{ij}^a \in \mathbb{Z}_{\geq 0}$, and $V_i$ denotes the representation of the Virasoro algebra $\mathrm{Vir}_c$ with conformal weight $h_i$, 
and similarly for $\bar{V}_{j}$ with $\bar{c} = c$.  

The proof proceeds in two steps: we first show that nonzero solutions to the boundary conformal condition exist only in the subspace $V_i \otimes \bar{V}_{j}$ with $h_i = h_j$. Next, we show that within each such subspace $V_i \otimes \bar{V}_j$,  
all solutions to the equation \eqref{bdy conformal} are $\mathbb{C}$-linearly dependent.  
After fixing the overall normalization, this implies that the solution is unique within each $V_i \otimes \bar{V}_j$.  

\subsection*{Step 1: Solution to Boundary Conformal Condition }

Suppose that there exists a nonzero solution in $V \otimes \bar{V}'$ with conformal weight $(h,h')$, which we denote by $\ishi{I}$. Then $\ishi{I}$ cannot be orthogonal to all $L_{-\{n_i\}}\ket{h} \coloneqq L_{-n_1}L_{-n_2}\cdots\ket{h}$'s, where $\ket{h}$ is the highest-weight state and $n_i\in\mathbb{Z}_{\geq 0}$. That is, there exist $\{n_i\}$ and $\{\bar{n}_j\}$ such that 
\begin{align}
    \bra{h} \otimes \bra{\bar{h}'}
    L_{\{n_i\}} \bar{L}_{\{\bar{n}_j\}}
    \ishi{I} \neq 0.
\end{align}
Here we use the Hermitian conjugate of chiral modes $L^\dagger_{-n}=L_n$.
Since $L_0\ishi{I}=\bar{L}_0\ishi{I}$, the state $\ishi{I}$ must be spinless. By the selection rule, the basis vector $L_{-\{n_i\}} \bar{L}_{-\{\bar{n}_j\}} \ket{h}\otimes \ket{\bar{h}'}$ must also be spinless, so that
\begin{align}
    h+\sum_i n_i = h'+\sum_j \bar{n}_j.
\end{align}

If $h>h'$, then $\sum_i n_i<\sum_j \bar{n}_j$. Using $\bar{L}_{\bar{n}_j}\ishi{I}=L_{-\bar{n}_j}\ishi{I}$, we have
\begin{align}
    0\neq
    \bra{h} \otimes \bra{\bar{h}'}
    L_{\{n_i\}} \bar{L}_{\{\bar{n}_j\}}
    \ishi{I}
    =
     \bra{h} \otimes \bra{\bar{h}'}
     L_{\{n_i\}} L_{-\{\bar{n}_j\}}
     \ishi{I}.
\end{align}
However,
\begin{align}
    \bra{h} \otimes \bra{\bar{h}'}
    L_{\{n_i\}} L_{-\{\bar{n}_j\}} 
    = 
    \bigl( L_{\{\bar{n}_j\}} L_{-\{n_i\}} \ket{h}\otimes\ket{\bar{h}'}\bigr)^\dagger
    = 0,
\end{align}
because $\sum_j\bar{n}_j-\sum_i n_i>0$ implies that $L_{\{\bar{n}_j\}} L_{-\{n_i\}} $ is a net annihilation operator.  
We thus reach a contradiction.

If $h<h'$, we obtain a similar contradiction from the fact that 
$\bar{L}_{\{n_i\}} \bar{L}_{-\{\bar{n}_j\}} \ket{h}\otimes\ket{\bar{h}'}=0$.  
Therefore, only the subspaces with $h=h'$ admit solutions to the conformal condition.

\subsection*{Step 2: Uniqueness in Subspace $V_h\otimes\bar{V}_h$}

For $h=h'$, we have $\sum_i n_i = \sum_j \bar{n}_j$. Thus the uniqueness of highest weight state implies
\begin{align}
    L_{\{\bar{n}_j\}} L_{-\{n_i\}} \ket{h}\otimes\ket{\bar{h}} \propto \ket{h}\otimes\ket{\bar{h}}.
\end{align}
Hence,
\begin{align}\label{nonzero overlap}
    \bra{h}\otimes\bra{\bar{h}}I\rangle\!\rangle \neq 0.
\end{align}
Suppose that both $\ishi{I}$ and $\ishi{I'}$ are nonzero solutions to equation \eqref{bdy conformal}.  
We define $\lambda \equiv \bra{h}\otimes\bra{\bar{h}}I\rangle\!\rangle \neq0$, $\lambda' \equiv \bra{h}\otimes\bra{\bar{h}}I'\rangle\!\rangle\neq0$.

Since equation \eqref{bdy conformal} is linear, the linear combination $\lambda' \ishi{I}-\lambda \ishi{I'}$ is also a solution, and it satisfies
\begin{align}
    \bra{h}\otimes\bra{\bar{h}}
    \left( \lambda'\ishi{I}-\lambda\ishi{I'} \right)=0.
\end{align}
But we have already shown that any nonzero solution to \eqref{bdy conformal} must satisfy \eqref{nonzero overlap}.  
Therefore, $\lambda'\ishi{I}-\lambda\ishi{I'}=0$, which means that within each spinless subspace $V \otimes \bar{V}'$,  
all solutions to \eqref{bdy conformal} are $\mathbb{C}$-linearly dependent. 

After fixing the conventional normalization, we conclude that the solutions to equation \eqref{bdy conformal} are spanned by the Ishibashi states.

\section{Compact Boson with it's SymTFT on Euclidean Lattice}\label{lattice}

In this section, we reformulate the content of Section \ref{sec: boson} in the language of Euclidean lattice models. In Section \ref{lattice1}, we review the modified Villain formulation of the compact boson \cite{Gorantla:2021svj, Choi:2021kmx}, and its Dirichlet and Neumann boundary conditions. The construction of the SymTFT and its topological boundary on the Euclidean lattice are discussed in Section \ref{lattice2}, in close parallel with Section \ref{sec: symtftbosonfield}. In Section \ref{lattice3}, we provide an alternative explanation for why the $\mathbb{Z}_p \times \mathbb{Z}_q$-preserving boundary does not admit a conformal physical corner $\widetilde{B}$ with the physical boundary $\widetilde{Q}$. This is a sanity check for the \textbf{Conformal condition} in the criteria.

\subsection{Modified Villain Formulation and its Known Boundary Conditions}\label{lattice1}
We first review the 2D Euclidean XY model in the modified Villain formulation on a closed (periodic boundary condition) two-dimensional square lattice following \cite{Gorantla:2021svj,Choi:2021kmx}:
\begin{align}\label{XY}
    \frac{R^2}{4\pi}\sum_{\text{link}}\left(\Delta X - 2\pi n\right)^2
    + i\sum_{\text{plaquette}}\widetilde{X}\,\Delta n.
\end{align}
Here $X(\hat{x},\hat{y})\in\mathbb{R}$ is defined on each site, $n^{(1)}\in\mathbb{Z}$ is defined on each link, and $\widetilde{X}\in\mathbb{R}$ is defined on each plaquette.  
For any $p$-form $a^{(p)}$, the lattice exterior derivative $\Delta a^{(p)}$ is a $(p+1)$-form given by the oriented sum of $a^{(p)}$ along the $p$-cells in the boundary of the $(p+1)$-cell.

The field $X$ is real-valued, but its integer part is gauged by the integer gauge field $n$:
\begin{align}
    X \to X + 2\pi k(\hat{x},\hat{y}), 
    \quad n \to n + \Delta k(\hat{x},\hat{y}),
    \quad k\in\mathbb{Z}.
\end{align}
Similarly, the shift
\begin{align}
    \widetilde{X} \to \widetilde{X} + 2\pi\widetilde{k}(\hat{x},\hat{y}),
    \quad \widetilde{k}\in\mathbb{Z},
\end{align}
leaves the action invariant, since $2\pi i\widetilde{k}\,\Delta n \in 2\pi i\mathbb{Z}$ does not change $e^{-S}$.

The winding number of a configuration is
\begin{align}
    \sum_{\gamma}(\Delta X - 2\pi n) 
    = -2\pi\sum_{\gamma} n.
\end{align}
A nonzero $\Delta n$ corresponds to vortices (or monopoles in higher dimensions, which we will encounter later), and the field $\widetilde{X}$ serves as a Lagrange multiplier enforcing the flatness condition.

This theory has a $U(1)$ momentum symmetry shifting $X$: $X\to X+\alpha$. The associated charge and current are
\begin{align}\label{mom lattice}
    Q(\widetilde{C}) = \sum_{\text{dual link}} \epsilon_{\mu\nu} J_\nu,\quad J_\mu = \frac{iR^2}{2\pi}\left(\Delta_\mu X - 2\pi n_\mu\right).
\end{align}
It also has a $U(1)$ winding global symmetry, under which $\widetilde{X}$ is charged: $\widetilde{X}\to\widetilde{X}+\widetilde{\alpha}$. This symmetry follows from the fact that $i\widetilde{\alpha}\sum \Delta n = 0$ on a closed lattice, but it may be broken on an open lattice. The corresponding charge and current are
\begin{align}\label{win lattice}
    \widetilde{Q}(C) = \sum_{\text{link}}\epsilon_{\mu\nu}\widetilde{J}_\nu
    ,\quad \widetilde{J}_\mu = \frac{\epsilon_{\mu\nu}}{2\pi}\left(\Delta_\nu X - 2\pi n_\nu\right).
\end{align}

Besides making the winding symmetry explicit, another important feature of the modified Villain formulation is its exact T-duality. We first ``sum by parts" the second term in the action.  
On a closed lattice, the boundary term vanishes.  
Then we apply the Poisson resummation formula to each link variable $n$.  
Up to an overall normalization of the partition function, the action becomes\footnote{The dual action also involves a term $\frac{i}{2\pi}\sum\Delta X\we\Delta\widetilde{X}$, which is zero on closed lattice.}
\begin{align}
    \frac{1}{4\pi R^2}
      \sum_{\text{dual link}}
      \left(\Delta \widetilde{X} - 2\pi \widetilde{n}\right)^2
      + i\sum_{\text{dual plaquette}} X\Delta \widetilde{n}.
\end{align}
In other words, the theory is dual to the same model, but defined on the dual lattice, with the radius $1/R$.

To go to the continuum limit, we sum over $\widetilde{X}$, which imposes the flatness on $n$. We then define a multi-valued variable:
\begin{align}
    \f(0,0)=X(0,0),\quad\Delta\f=\Delta X-2\pi n.
\end{align}
Then we goes back to the continuous compact $\frac{R^2}{4\pi}\partial_\mu\f\partial^\mu\f$.

We now discuss how Dirichlet and Neumann boundary conditions are realized in the lattice Villain formulation and how they transform under T-duality.

\begin{figure}
\centering
\begin{tikzpicture}[scale=0.8]
\begin{scope}
    \draw (0,1)--(6,1);
    \draw (0,2)--(6,2);
    \draw (0,3)--(6,3);
    \draw (0,4)--(6,4);
    \draw (1,0)--(1,5);
    \draw (2,0)--(2,5);
    \draw (3,0)--(3,5);
    \draw (4,0)--(4,5);
    \draw (5,0)--(5,5);
    \draw[gray!30] (0.5,0.5)--(5.5,0.5);
    \draw[gray!30] (0.5,1.5)--(5.5,1.5);
    \draw[gray!30] (0.5,2.5)--(5.5,2.5);
    \draw[gray!30] (0.5,3.5)--(5.5,3.5);
    \draw[gray!30] (0.5,4.5)--(5.5,4.5);
    \draw[gray!30] (0.5,0.5)--(0.5,4.5);
    \draw[gray!30] (1.5,0.5)--(1.5,4.5);
    \draw[gray!30] (2.5,0.5)--(2.5,4.5);
    \draw[gray!30] (3.5,0.5)--(3.5,4.5);
    \draw[gray!30] (4.5,0.5)--(4.5,4.5);
    \draw[gray!30] (5.5,0.5)--(5.5,4.5);
    \node[circle, fill=black, inner sep=1pt] at (2,1) {};
    \node at (1.6,0.6){\footnotesize{$X$}};
    \draw[thick, decoration = {markings, mark=at position 0.6 with {\arrow[scale=1.0]{stealth}}}, postaction=decorate] (3,1) -- (4,1);
    \draw[thick, decoration = {markings, mark=at position 0.6 with {\arrow[scale=1.0]{stealth}}}, postaction=decorate] (3,1) -- (3,2);
    \node[circle, fill=gray, inner sep=1pt] at (4.5,2.5) {};
    \node[gray] at (4.3,2.5){\footnotesize{$\widetilde{X}$}};
    \draw[fill=gray!20,opacity=0.5] (1,3) -- (2,3) -- (2,4) -- (1,4) -- cycle;
    \node[gray] at(1.5,3.5){$\Delta n$};
    \node at(1.5,5.5){\footnotesize{$n_{\mu}\big|=0$}};
    \node at(4,5.5){\footnotesize{$X\big|=\theta$}};
    \node[circle, fill=gray, inner sep=1pt] at (4,5) {};
\end{scope}
\begin{scope} [xshift=8cm]
    \draw (0,0)--(6,0);
    \draw (0,1)--(6,1);
    \draw (0,2)--(6,2);
    \draw (0,3)--(6,3);
    \draw (0,4)--(6,4);
    \draw (0,5)--(6,5);
    \draw (0,0)--(0,5);
    \draw (1,0)--(1,5);
    \draw (2,0)--(2,5);
    \draw (3,0)--(3,5);
    \draw (4,0)--(4,5);
    \draw (5,0)--(5,5);
    \draw (6,0)--(6,5);
    \draw[gray!30] (-0.5,0.5)--(6.5,0.5);
    \draw[gray!30] (-0.5,1.5)--(6.5,1.5);
    \draw[gray!30] (-0.5,2.5)--(6.5,2.5);
    \draw[gray!30] (-0.5,3.5)--(6.5,3.5);
    \draw[gray!30] (-0.5,4.5)--(6.5,4.5);
    \draw[gray!30] (0.5,-0.5)--(0.5,5.5);
    \draw[gray!30] (1.5,-0.5)--(1.5,5.5);
    \draw[gray!30] (2.5,-0.5)--(2.5,5.5);
    \draw[gray!30] (3.5,-0.5)--(3.5,5.5);
    \draw[gray!30] (4.5,-0.5)--(4.5,5.5);
    \draw[gray!30] (5.5,-0.5)--(5.5,5.5);
        \node[circle, fill=black, inner sep=1pt] at (2,1) {};
    \node at (1.6,0.6){\footnotesize{$X$}};
    \draw[thick, decoration = {markings, mark=at position 0.6 with {\arrow[scale=1.0]{stealth}}}, postaction=decorate] (3,1) -- (3,2);
    \draw[thick, decoration = {markings, mark=at position 0.6 with {\arrow[scale=1.0]{stealth}}}, postaction=decorate] (3,1) -- (4,1);
    \node[circle, fill=gray, inner sep=1pt] at (3.5,4.5) {};
    \node[gray] at (3.3,4.5){\footnotesize{$\widetilde{X}$}};
    \draw[fill=gray!20,opacity=0.5] (1,4) -- (2,4) -- (2,5) -- (1,5) -- cycle;
    \node[gray] at(1.5,4.5){$\Delta n$};
\end{scope}
\end{tikzpicture}
    \caption{Left: rough boundary and Dirichlet boundary condition; Right: smooth boundary and Neumann boundary condition}
    \label{fig:XY bdy}
\end{figure}
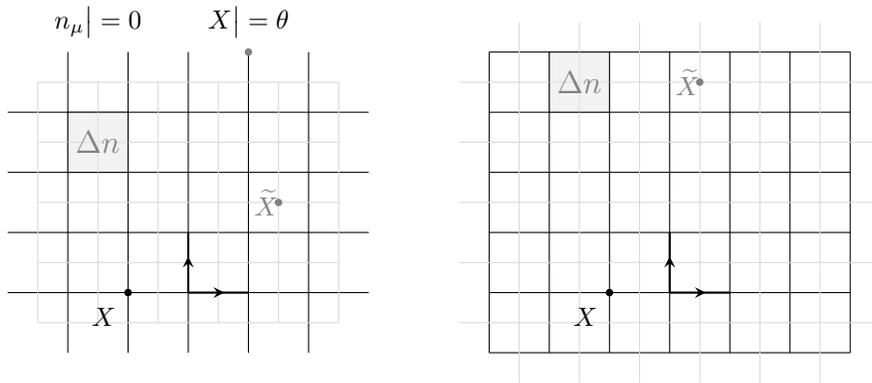

We impose Dirichlet boundary conditions by fixing $\Delta_{\tau}X-2\pi n_\tau|=0$. We can choose a special gauge:
\begin{align}
    n|= 0,
    \quad
    X| = \theta,
\end{align}
where $\theta$ is a fixed boundary value of the field. This can be thought of as a ``rough boundary" of the lattice geometry: boundary links are empity, meaning that the $n$'s valued on it are fixed to be zero, and these links make no contributions to the kinetic term.

Now we check its consistency with T-duality. Since $n|=0$, there are no boundary contributions to the ``sum by parts", and hence we can still apply the resummation formula to each link integer $n$. Note that in the presence of this boundary, the dual action is accompanied by an additional boundary term:
\begin{align}
    S_R[X, n]\xrightarrow{\,\text{T-dual}\,} \widetilde{S}_{\frac{1}{R}}[\widetilde{X}, \widetilde{n}]+i\theta\sum_{\text{bdy link}}\widetilde{n}.
\end{align}
Under T-duality, the original lattice with ``rough boundary" is mapped to the dual lattice with ``smooth boundary".

A naive guess for Neumann boundary conditions might be to fix
$\widetilde{X}|_{\text{bdy plaquette}} = \widetilde{\theta}$.
However, in this formulation $\widetilde{X}$ is a Lagrange multiplier enforcing vortex constraints, so this direct prescription is not meaningful. Instead, we impose Neumann boundary conditions by leaving $X$ and $n$ unconstrained along the boundary\footnote{The condition $\partial_n\f|=0$ in the continuum formulation becomes $\Delta_xX-2\pi n_x|=0$}, and adding a $\widetilde{\theta}$-dependent boundary term to the action:
\begin{align}\label{free bdy}
    S_R[X,n]+i\widetilde{\theta}\sum_{\text{bdy link}}n.
\end{align}
We refer to it as the ``smooth boundary" of the lattice. We note that the dual lattice has ``rough boundary", After performing the T-duality, the theory is mapped to the dual lattice with boundary condition $\widetilde{X}|=\widetilde{\theta}$ and $\widetilde{n}|=0$, which is the Dirichlet boundary condition for the dual theory,
\begin{align}
    S_R[X,n]+i\widetilde{\theta}\sum_{\text{bdy link}}n\xrightarrow{\,\text{T-dual}\,} \widetilde{S}_{\frac{1}{R}}[\widetilde{X}, \widetilde{n}].
\end{align}
In summary, T-duality exchanges Dirichlet and Neumann boundary conditions on the lattice:

\begin{table}
\centering
\begin{tabular}{|c|c|}
\hline
Original theory & Dual theory \\
\hline
Dirichlet (rough boundary) & Neumann (smooth boundary) \\
\hline
Neumann (smooth boundary) & Dirichlet (rough boundary) \\
\hline
\end{tabular}
\caption{Boundary conditions and T duality}
\end{table}

Thus, the duality is preserved in the open-lattice case, but boundary types and lattice geometry are exchanged.

\subsection{SymTFT, Topological Boundaries and Physical Boundary}\label{lattice2}
The SymTFT for the $U(1)\times U(1)$ symmetry with a mixed anomaly is described by two real-valued gauge fields in \eqref{S3}; its lattice version is similar.

First, we define the corresponding TQFT on a 3d cubic lattice without boundary ($T^3$). We take a 3d cubic lattice, assign $\omega\in\mathbb{R}$ to each link and $\widetilde{\omega}\in\mathbb{R}$ to each dual link. The real gauge transformations are
\begin{align}
    \omega \to \omega + \Delta \eta,\qquad
    \widetilde{\omega} \to \widetilde{\eta} + \Delta \widetilde{\alpha},
\end{align}
where $\eta\in\mathbb{R}$ is defined on each site and $\widetilde{\eta}\in\mathbb{R}$ is defined on each dual site.

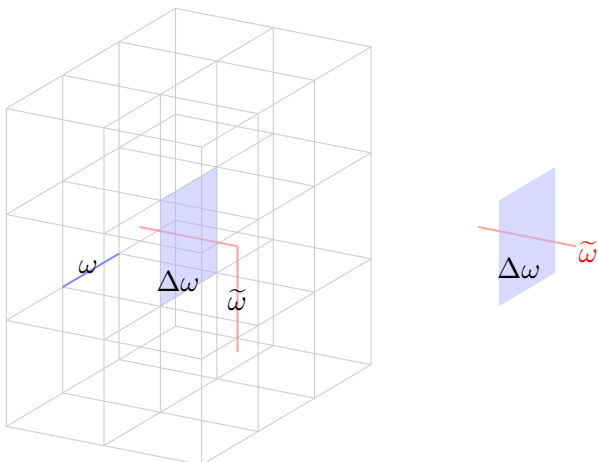
\begin{figure}
    \centering
    \tdplotsetmaincoords{70}{120}
	\begin{tikzpicture}[scale=1.5,tdplot_main_coords,
		edge/.style={gray!40,thin},
		interface/.style={fill=blue!30,opacity=0.5},
		specialline/.style={thick},
		axislabel/.style={font=\small}
		]
        \begin{scope}
		\def\sizex{3}
		\def\sizey{2}
		\def\sizez{3}
		
		\foreach \x in {0,...,\sizex}
		\foreach \y in {0,...,\sizey}
		\foreach \z in {0,...,\sizez} {
			\ifnum\x<\sizex
			\draw[edge] (\x,\y,\z) -- (\x+1,\y,\z);
			\fi
			\ifnum\y<\sizey
			\draw[edge] (\x,\y,\z) -- (\x,\y+1,\z);
			\fi
			\ifnum\z<\sizez
			\draw[edge] (\x,\y,\z) -- (\x,\y,\z+1);
			\fi
		}

		\coordinate (center1) at (1.5,1.5,1.5); 
		\coordinate (center2) at (1.5,0.5,1.5); 
		\draw[red!50,opacity=0.7,specialline] (center1) -- (center2);
		
		\coordinate (bottomCenterLeft) at (1.5,1.5,0.5);
		\coordinate (topCenterLeft) at (1.5,1.5,1.5);
		\draw[red!50,opacity=0.7,specialline] (bottomCenterLeft) -- (topCenterLeft);
		\coordinate (omegaStart) at (1,0,1);
		\coordinate (omegaEnd) at (2,0,1);
		\draw[blue!80,opacity=0.5,specialline] (omegaStart) -- (omegaEnd);

        \fill[interface] (1,1,1) -- (2,1,1) -- (2,1,2) -- (1,1,2) -- cycle;
        
		\node[axislabel] at (1.6,0.95,1.1) {$\Delta\omega$};
        \node[axislabel] at (1.5,1.5,1.0) {$\widetilde{\omega}$};
        \node[axislabel] at (1.2,-0.2,0.9) {$\omega$};
     
        \end{scope}

        \begin{scope}[xshift=3cm]
            \def\faceX{1}
            \def\faceY{1} 
            \def\faceZ{1}
            \def\faceSize{1}
            \coordinate (topCenter) at (\faceX+0.5,\faceY+0.5,\faceZ+0.5);
            \coordinate (bottomCenter) at (\faceX+0.5,\faceY-0.5,\faceZ+0.5);
            \draw[red!50,opacity=0.7,specialline] (bottomCenter) -- (topCenter);
            \fill[interface] 
                (\faceX,\faceY,\faceZ) -- 
                (\faceX+\faceSize,\faceY,\faceZ) -- 
                (\faceX+\faceSize,\faceY,\faceZ+\faceSize) -- 
                (\faceX,\faceY,\faceZ+\faceSize) -- cycle;

            \node[axislabel] at (\faceX+0.3,\faceY-0.2,\faceZ+0.1) {$\Delta\omega$};
            \node[axislabel,red] at (\faceX+0.8,\faceY+0.8,\faceZ+0.6) {$\widetilde{\omega}$};
        \end{scope}
	\end{tikzpicture}
    \caption{Left: The cubic lattice where the 3d TQFT is defined on. Right: One term $\widetilde{\om}\Delta\om$ in the action \eqref{S3 lattice}}
\end{figure}

The action is
\begin{align}\label{S3 lattice}
    S_3=\frac{i}{2\pi}\sum_{\text{plaquette}} \widetilde{\omega}\,\Delta\omega.
\end{align}
Here $\widetilde{\omega}$ serves as a Lagrange multiplier enforcing the flatness condition of $\omega$, i.e. $\Delta\omega=0$ on every plaquette. After summation by parts, $\omega$ plays the role of Lagrangian multiplier enforcing the flatness of $\widetilde{\omega}$.

We now discuss the topological boundary conditions of this TQFT. 
\paragraph{Dirichlet boundary $\mathcal{B}_0$}
Similar to the continuous case, one option is to set
\begin{align}
   \text{Dirichlet:}\quad \omega\big|=0,
\end{align}
which can be thought of as a ``rough boundary'' of the cubic lattice. 
\begin{align}
    L_{(x,0)}\coloneqq\exp\left(ix\sum_{\gamma\subset\text{bdy links}}\omega\right),\quad x\in\mathbb{R},
\end{align}
becomes the identity on the boundary. It can be explicitly checked that the theory defined on the open lattice is gauge invariant.

\paragraph{Neumann boundary $\mathcal{B}_\infty$}
We allow the boundary $\omega$ to fluctuate freely. For the same reason explained above \eqref{free bdy}, we cannot directly set $\widetilde{\omega}\big|$ to zero on the boundary. Instead, we introduce dual links orthogonal to the boundary plaquettes and place $\widetilde{\omega}_\perp$ on those dual links as Lagrange multipliers to enforce the flatness of the boundary $\omega$. This corresponds to a ``smooth boundary" for the original lattice (and a rough boundary for the dual lattice).

\begin{figure}
    \centering
    \tdplotsetmaincoords{70}{120}
	\begin{tikzpicture}[scale=1.5,tdplot_main_coords,
		edge/.style={gray!40,thin},
		specialline/.style={red!50,opacity=0.7,thick},
		face/.style={gray!20,fill opacity=0.6}
		]
        \begin{scope}[xshift=0cm]
        	\def\sizex{3}
		\def\sizey{2}
		\def\sizez{3}
		\foreach \x in {0,...,\sizex}
		\foreach \y in {0,...,\sizey}
		\foreach \z in {0,...,\sizez} {
			\ifnum\x<\sizex
			\draw[edge] (\x,\y,\z) -- (\x+1,\y,\z);
			\fi
			\ifnum\y<\sizey
			\draw[edge] (\x,\y,\z) -- (\x,\y+1,\z);
			\fi
			\ifnum\z<\sizez
			\draw[edge] (\x,\y,\z) -- (\x,\y,\z+1);
			\fi
		}
		\foreach \x in {0,1,2}  
		\foreach \z in {0,1,2} {
			\fill[face] (\x,2,\z) -- (\x+1,2,\z) -- (\x+1,2,\z+1) -- (\x,2,\z+1) -- cycle;
		}
		\node[darkgray,font=\footnotesize] at (1.5,2,1) {$\omega\big|_\partial=0$};
		\draw[orange!50,line width=4pt] (3,2,2) -- (0,2,2);
		\node[orange,font=\footnotesize] at (2.5,2.5,2.8) {$\sum_{\gamma}\omega=0$};
        \end{scope}
        
        \begin{scope}[xshift=4cm]
        \def\sizex{3}
		\def\sizey{2}
		\def\sizez{3}
		\foreach \x in {0,...,\sizex}
		\foreach \y in {0,...,\sizey}
		\foreach \z in {0,...,\sizez} {
			\ifnum\x<\sizex
			\draw[edge] (\x,\y,\z) -- (\x+1,\y,\z);
			\fi
			\ifnum\y<\sizey
			\draw[edge] (\x,\y,\z) -- (\x,\y+1,\z);
			\fi
			\ifnum\z<\sizez
			\draw[edge] (\x,\y,\z) -- (\x,\y,\z+1);
			\fi
		}
		\draw[orange!50,line width=4pt] (1.5,1.5,1.5) -- (1.5,2.5,1.5);
		\draw[orange!50,line width=4pt] (1.5,1.5,1.5) -- (1.5,1.5,0.5);
		\draw[orange!50,line width=4pt] (1.5,1.5,0.5) -- (1.5,2.5,0.5);
		\draw[green,line width=2pt] (1,2,1) -- (2,2,1);
		\foreach \x in {0,1,2}  
		\foreach \z in {0,1,2} {  
			\pgfmathsetmacro{\cx}{\x+0.5}
			\pgfmathsetmacro{\cy}{1.5}
			\pgfmathsetmacro{\cz}{\z+0.5}
			\fill[face] (\x,2,\z) -- (\x+1,2,\z) -- (\x+1,2,\z+1) -- (\x,2,\z+1) -- cycle;
			\draw[specialline] (\cx,\cy,\cz) -- (\cx,\cy+1,\cz);
		}
		\node[red] at (1.5,2.0,2.5) {$\widetilde{\omega}_\perp$};
		\draw[orange!50,line width=4pt] (1.5,1.5,1.5) -- (1.5,2.5,1.5);
		\draw[orange!50,line width=4pt] (1.5,1.5,1.5) -- (1.5,1.5,0.5);
		\draw[orange!50,line width=4pt] (1.5,1.5,0.5) -- (1.5,2.5,0.5);
		\draw[green,line width=2pt] (1,2,1) -- (2,2,1);
		\node[darkgray] at (1.5,2,1) {$\omega\big|_\partial$};
		\node[orange,font=\footnotesize] at (1.5,2.5,0.3) {$\sum_{\widetilde{\gamma}}\widetilde{\omega}=0$};
        \end{scope}

        \begin{scope}[xshift=8cm]
		\def\sizex{3}
		\def\sizey{2}
		\def\sizez{3}
		\foreach \x in {0,...,\sizex}
		\foreach \y in {0,...,\sizey}
		\foreach \z in {0,...,\sizez} {
			\ifnum\x<\sizex
			\draw[edge] (\x,\y,\z) -- (\x+1,\y,\z);
			\fi
			\ifnum\y<\sizey
			\draw[edge] (\x,\y,\z) -- (\x,\y+1,\z);
			\fi
			\ifnum\z<\sizez
			\draw[edge] (\x,\y,\z) -- (\x,\y,\z+1);
			\fi
		}
		\draw[orange!50,line width=4pt] (1.5,1.5,1.5) -- (1.5,2.5,1.5);
		\draw[orange!50,line width=4pt] (1.5,1.5,1.5) -- (1.5,1.5,0.5);
		\draw[orange!50,line width=4pt] (1.5,1.5,0.5) -- (1.5,2.5,0.5);
		\draw[green,line width=2pt] (1,2,1) -- (2,2,1);
		\foreach \x in {0,1,2}  
		\foreach \z in {0,1,2} {  
			\pgfmathsetmacro{\cx}{\x+0.5}
			\pgfmathsetmacro{\cy}{1.5}
			\pgfmathsetmacro{\cz}{\z+0.5}
			\fill[face] (\x,2,\z) -- (\x+1,2,\z) -- (\x+1,2,\z+1) -- (\x,2,\z+1) -- cycle;
			\draw[specialline] (\cx,\cy,\cz) -- (\cx,\cy+1,\cz);
		}
		\node[red] at (1.5,2.0,2.5) {$\widetilde{\omega}_\perp$};
		\draw[orange!50,line width=4pt] (1.5,1.5,1.5) -- (1.5,2.5,1.5);
		\draw[orange!50,line width=4pt] (1.5,1.5,1.5) -- (1.5,1.5,0.5);
		\draw[orange!50,line width=4pt] (1.5,1.5,0.5) -- (1.5,2.5,0.5);
		\draw[green,line width=2pt] (1,2,1) -- (2,2,1);
		\node[darkgray,font=\footnotesize] at (1.5,2.5,1) {$\omega\big|_\partial\in 2\pi r \mathbb{Z}$};
		\node[orange,font=\footnotesize] at (1.5,2.5,0.3) {$\sum_{\widetilde{\gamma}}\widetilde{\omega}\in\frac{2\pi}{r}\mathbb{Z}$};
        \end{scope}
	\end{tikzpicture}
    \caption{Left: Dirichlet boundary $\mathcal{B}_0$. Middle: Nuemann boundary $\mathcal{B}_\infty$. Right: boundary $\mathcal{B}_r$.}
    \label{topo bdy lattice}
\end{figure}
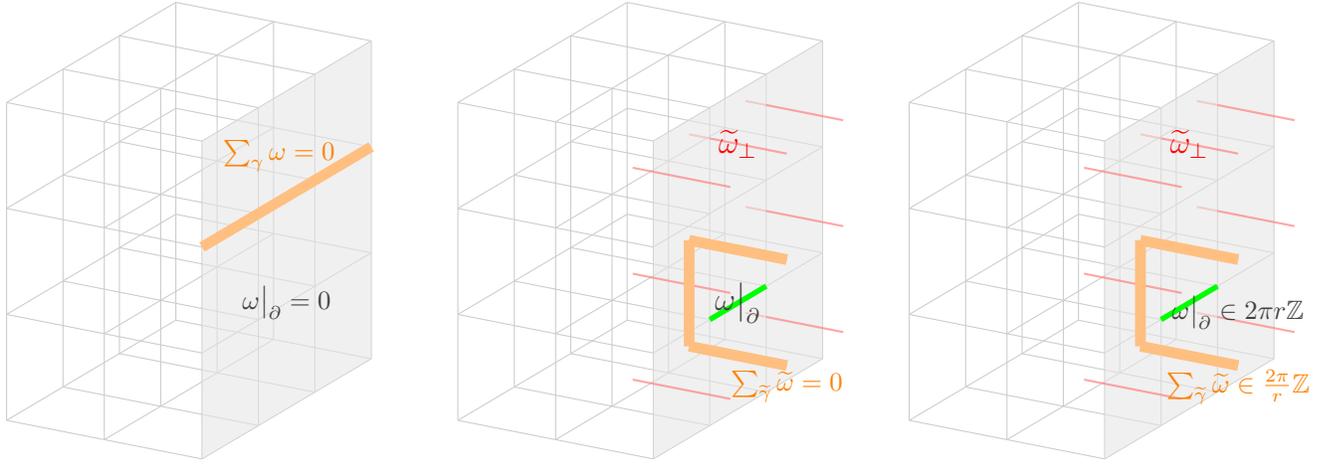

We claim that the line operator
\begin{align}
    L_{(0,y)}\coloneqq\exp\left(iy\sum_{\widetilde{\gamma}\subset\text{dual bdy links}}\widetilde{\om}\right),\quad y\in\mathbb{R}
\end{align}
 becomes the identity on this boundary for the following reason. After ``summing by parts", the action is written as a sum over dual plaquettes of the dual lattice with a rough boundary: boundary plaquettes (whose normals are parallel to the boundary) contain only three links (the orange highlighted line in the middle of Figure \ref{topo bdy lattice}), and the $\omega$-sum (the green highlighted line in Figure \ref{topo bdy lattice}) enforces that the corresponding $\widetilde{\omega}$ on those dual links vanish. 

\paragraph{Boundary $\mathcal{B}_r$}A further class of topological boundary conditions fixes the value of $\omega$ on the boundary only up to an integer multiple of $2\pi r$:
\begin{align}\label{r bdy}
    \omega\big|\in 2\pi r\,\mathbb{Z},
\end{align}
where $r$ is any fixed real number.  

To implement the flatness condition for the boundary $\omega$, one also needs to introduce dual links orthogonal to the boundary plaquettes and define $\widetilde{\omega}_{\perp}$ on them. We then have the following line operators identified with the identity:
\begin{align}
    L_{(\frac{1}{r}, 0)}\coloneqq\exp\left(\frac{i}{r}\sum_{\gamma}\om\right)=1,\quad\text{and}\quad L_{(0,r)}\coloneqq\exp\left(ir\sum_{\widetilde{\gamma}}\widetilde{\om}\right)=1.
\end{align}
The formal equality $L_{\left(\frac{1}{r}, 0\right)} = 1$ directly comes from the boundary condition \eqref{r bdy}. The latter equality $L_{(0,r)}=1$  (the orange highlighted line in the right of Figure \ref{topo bdy lattice})is obtained by summing over the integers in \eqref{r bdy} defined on the boundary links (the green highlighted line in Figure \ref{topo bdy lattice}) intersecting $\widetilde{\gamma}$. 

Now we discuss the physical boundary $\widetilde{Q}$. The kinetic term is similar to the continuum version \eqref{S2phy}, but we still need to introduce Lagrange multipliers $\widetilde{X}$ to enforce the flatness condition of $\omega$ on the physical boundary. The action is 
\begin{align}
    S_2^{\text{phy}}=\frac{R^2}{4\pi}\sum_{\text{link}\subset\mathcal{B}^{\text{phy}}}\left(\Delta X-\om\right)^2+\frac{i}{2\pi}\sum_{\text{plaquette}\subset\mathcal{B}^{\text{phy}}}\widetilde{X}\Delta\om.
\end{align}
The gauge transformations are extended to the physical boundary by $X\to X+\eta$ and $\om\to\om+\eta$, with $\eta\in\mathbb{R}$. The SymTFT for the compact boson on the Euclidean lattice without boundary is completely specified by $(\mathcal{B}_1, S_3, S_2^{\text{phy}})$. After dimensional reduction, the link variables $\omega$, defined on links of a 2d square lattice, are enforced to take values in $2\pi \mathbb{Z}$, and we recover the 2d action \eqref{XY}.

\subsection{Conformal Boundary Conditions}\label{lattice3}
We have the correspondence between symmetry operators in 2d compact boson \eqref{mom lattice}, \eqref{win lattice} and lines operators supported on the symmetry boundary
\begin{align}
    \text{momentum:}\quad&\exp\left(i\beta\sum_{\widetilde{\gamma}} \frac{iR^2}{2\pi}\epsilon_{\mu\nu}(\Delta_\nu X-2\pi n_\nu) \right) \longleftrightarrow \exp\left( i\frac{\beta}{2\pi}\sum_{\widetilde{\gamma}}\widetilde{\om}\right)\\
    \text{winding:}\quad&\exp\left(i\al\sum_{\gamma}\frac{1}{2\pi}(\Delta X_\mu-2\pi n_\mu)\right)\longleftrightarrow\exp\left(i\frac{\al}{2\pi}\sum_\gamma\om\right).
\end{align}
The off-diagonal component of stress tensor is proportional to $(\Delta_\tau X-2\pi n_\tau)(\Delta_x X-2\pi n_x)$. For conformal boundary conditions, we have the constraint that $T_{\tau x}|=0$, which means that on the physical corner $\widetilde{B}$, either $\om$ or $\widetilde{\om}$ is supposed to be zero. 

The condition is satisfied if we choose  the Dirichlet $\mathcal{B}_0$ or Neumann $\mathcal{B}_\infty$ boundary as $\mathcal{B}^{\text{bdy}}$. But for the $\mathcal{B}_r$ boundary condition, $\om\in2\pi r\mathbb{Z}$ and $\widetilde{\om}\in\frac{2\pi}{r} \mathbb{Z}$ on the physical corner, making it not conformal. The discrete values of $\om$ are preserved under renormalization flow, making it not a conformal boundary condition for compact boson CFT.

\section{Proof: Cardy States are \emph{All} Simple Boundary States of a Diagonal Minimal Model}\label{cardy}

In this section, we prove that in a unitary diagonal minimal model, any conformal boundary state is a superposition of Cardy states with non-negative integer coefficients. In other words, all simple boundary states of a diagonal minimal model are Cardy states. This result is well-known \cite{Fuchs:2002cm}, but we include an explicit proof for completeness.  This statement serves as a benchmark for the discussion in Section \ref{sec: min model}.

Using the results from Section~\ref{ishibashi}, we denote a general boundary state as 
\begin{align}
    \ket{B_\alpha} = \sum_i C_{B_\alpha i}\, \ket{i}\!\rangle,
\end{align}
where $\ket{i}\!\rangle$ is the Ishibashi state in the subspace $V_i \otimes \bar{V}_i$. With the conventional normalization of Ishibashi states, 
\begin{align}
    \langle\!\bra{i} e^{-H_{\text{cl}}/\delta} \ket{j}\!\rangle = \delta_{ij}\, \chi_i(e^{-4\pi/\delta}),
\end{align}
and using the $S$-matrix of a diagonal minimal model, the Cardy condition 
\begin{align}
    \operatorname{Tr}_{\mathcal{H}_{B_\alpha B_\beta}} e^{-H_{\text{op}}\delta}
    = \bra{B_\alpha} e^{-H_{\text{cl}}/\delta} \ket{B_\beta}, 
    \qquad 
    \mathcal{H}_{B_\alpha B_\beta} = \bigoplus_j n_{B_\alpha B_\beta}^j\, V_j
\end{align}
can be rewritten as
\begin{align}
    \sum_i C_{B_\alpha i}^*\, C_{B_\beta i}\, S_{ij} = n_{B_\alpha B_\beta}^j \in \mathbb{Z}_{\geq 0},
\end{align}
for each representation $V_i \otimes \bar{V}_i$.

Cardy found a set of solutions to these equations using the Verlinde formula:
\begin{align}
    C_{B_i j} = \frac{S_{ij}}{\sqrt{S_{1j}}}.
\end{align}
The label $\alpha$ of the boundary states $\{\ket{B_\alpha}\}$ corresponds to the label of primary fields.  
We now show that any boundary state satisfying the Cardy conditions with the Cardy states must be a superposition of Cardy states with non-negative integer coefficients.  
In other words, the Cardy states constitute \emph{all} the simple boundary states of a diagonal minimal model.

Suppose 
\begin{align}
    \ket{X} = \sum_i X_i\, \ket{i}\!\rangle
\end{align}
is a boundary state. Consider the Cardy condition between $\ket{X}$ and the Cardy state associated with the identity operator $1$: 
\begin{align}
    \ket{B_1} = \sum_i \sqrt{S_{1i}}\, \ket{i}\!\rangle.
\end{align}
We obtain
\begin{align}
    \sum_j S_{cj} \left( \sqrt{S_{1j}}\, X_j \right) = m_c \in \mathbb{Z}_{\ge 0}.
\end{align}

Define the vectors 
\begin{align}
    \vec{x} = \sum_j \sqrt{S_{1j}}\, X_j \, \vec{e}_j,
    \qquad 
    \vec{m} = \sum_j m_j\, \vec{e}_j,
\end{align}
where $\vec{e}_j = (0, \dots, 1, \dots, 0)^T$ is the standard basis vector.  
Then the above condition can be written compactly as
\begin{align}
    S \cdot \vec{x} = \vec{m}.
\end{align}
For a diagonal minimal model, we have $S^2 = 1$, hence
\begin{align}
    \vec{x} = S \cdot \vec{m} 
    = \sum_j m_j\, (S \cdot \vec{e}_j) 
    = \sum_{j,i} m_j\, S_{ji}\, \vec{e}_i.
\end{align}
Comparing with $\vec{x} = \sum_i \sqrt{S_{1i}}\, X_i\, \vec{e}_i$, we find
\begin{align}
    X_i = \sum_j m_j \frac{S_{ji}}{\sqrt{S_{1i}}},
\end{align}
or equivalently,
\begin{align}
    \ket{X} = \sum_i m_i\, \ket{B_i}.
\end{align}

\section{Proof: $\mathcal{D}\simeq\operatorname{Rep}(A_4)$ as Fusion Category}\label{near group}
This appendix consists of two parts. In Section \ref{near gp 1}, we collect several mathematical theorems from the literature, which are then used in Section \ref{near gp 2} to prove the statement \eqref{D rep} in Section \ref{sec: su3 wzw}.

\subsection{Some Theorems on Braided Near-group Categories}\label{near gp 1}

A fusion category $\mathcal{C}$ is called \emph{near-group} if \cite{Evans:2012ta,Thornton:2012GeneralizedNearGroup,Schopieray:2021Nondegenerate}:
\begin{enumerate}
    \item Its simple objects are labeled by the elements of a finite group $G=\{g,h,\dots\}$ and a single non-invertible object $X$:
    \begin{align}
        \operatorname{Irr}(\mathcal{C}) = \{g,h,\dots,X\}.
    \end{align}
    \item The fusion rules are
    \begin{align}
        g\otimes h = gh, \quad g\otimes X = X, \quad X^2 = \bigoplus_{g\in G} g \oplus k X = g\oplus h\oplus\cdots \oplus kX,
    \end{align}
    for an integer $k\geq 0$, where $gh$ is the group multiplication.
\end{enumerate}
We denote such a category as $(G,k)$\footnote{We hope readers do not get confused with the notation in Section \ref{sec: remark M(6,5)}, where $(A_1,k)$ denotes the category of Verlinde lines in $SU(2)_k$ WZW model.}. The Tambara-Yamagami fusion category $\mathrm{TY}(G)$ is a special case $(G,0)$. The $\mathrm{Rep}(S_3)$ encountered in Section \ref{sec: min model} is $(\bZ_2,1)$. The subcategory $\mathcal{D}$ in Section \ref{sec: su3 wzw} is also near-group of type $(\mathbb{Z}_3,2)$.

The braiding in a braided fusion category is called \emph{symmetric} if the double braiding of any two objects $a$ and $b$ is trivial:
\begin{align}
    R^{ba}R^{ab}=id: a\otimes b \to a\otimes b.
\end{align}

If one requires a near-group category to have a \emph{non-symmetric} braiding, the possibilities are very limited, this was proven in \cite[Theorem III.4.6]{Thornton:2012GeneralizedNearGroup}

\paragraph{Theorem III.4.6 \cite{Thornton:2012GeneralizedNearGroup}:}
Non-symmetrically braided near-group fusion categories with non-trivial $G$ and $k\neq 0$ have been classified, up to braided tensor equivalence, as follows:
\begin{itemize}
    \item $G = \mathbb{Z}_2$, $k=1$: two inequivalent braided fusion categories.
    \item $G = \mathbb{Z}_3$, $k=2$: a unique braided fusion category.
\end{itemize}
These three categories are explicitly realized in \cite[Example 5.0.2]{Schopieray:2021Nondegenerate}.
\begin{itemize}
    \item The two $(\mathbb{Z}_2,1)$ categories appear as braided fusion subcategories of the untwisted quantum double $\mathfrak{D}(S_3)$, generated by $([(123)], \omega)$ and $([(123)], \omega^*)$ respectively, where $[123]$ is the conjugacy class of order 3 in $S_3$, and $\omega$, $\omega^*$ are the two non-trivial 1-dimensional representations of centralizer $Z_{(123)} = \mathbb{Z}_3$.
    \item The $(\mathbb{Z}_3,2)$ category is a braided fusion subcategory of the untwisted quantum double $\mathfrak{D}(A_4)$, generated by $([(12)(34)], \pi)$, where $\pi$ is a representation of centralizer $Z_{(12)(34)} = \mathbb{Z}_2 \times \mathbb{Z}_2$ with character $\chi((12)(34))=-1$. \footnote{There are two different simple objects in $\mathfrak{D}(S_3)$ with these properties, but they generate the same category, as seen from the sub-$S$ and $T$ matrices.}
\end{itemize}

We now explicitly describe the $(\mathbb{Z}_3,2)$ braided fusion subcategory, which will be compared to $\mathcal{D}$ in Section \ref{sec: su3 wzw}. We started with $\mathfrak{D}(A_4)$ where $(\mathbb{Z}_3,2)$ is included in the theorem above. There are 14 simple objects in $\mathfrak{D}(A_4)$:
\begin{align}
\begin{array}{ccccccc}
    A & B & C & D & E & F & G \\
    ([e],1) & ([e],1') & ([e],1'') & ([e],3) & ([C_2],1) & ([C_2],1') & ([C_2],1'') \\[2mm]
    H & I & J & K & L & M & N \\
    ([C_2],1''') & ([C_3],1) & ([C_3],\omega) & ([C_3],\omega^*) & ([C_3^*],1) & ([C_3^*],\omega) & ([C_3^*],\omega^*)
\end{array}
\end{align}
with
\begin{itemize}
    \item $[e] = \{e\}$, $Z_e = A_4$;
    \item $[C_2] = \{(12)(34),(13)(24),(14)(23)\}$, $Z_{C_2} = \mathbb{Z}_2 \times \mathbb{Z}_2$;
    \item $[C_3] = \{(123),(124),(134)\}$, $Z_{C_3} = \mathbb{Z}_3$;
    \item $[C_3^*] = \{(132),(142),(143)\}$, $Z_{C_3^*} = \mathbb{Z}_3$.
\end{itemize}
The $S$ and $T$ matrices are given by \footnote{see, for example, \url{https://www.cpt.univ-mrs.fr/~coque/quantumdoubles/comments.html}.}:
\begin{equation}
S=\frac{1}{12}
\setlength{\arraycolsep}{3pt}
\renewcommand{\arraystretch}{1.1}
\left(
\begin{array}{cccccccc|cccccc}
1 & 1 & 1 & 3 & 3 & 3 & 3 & 3 & 4 & 4 & 4 & 4 & 4 & 4\\
1 & 1 & 1 & 3 & 3 & 3 & 3 & 3 & \alpha & \alpha & \alpha & \alpha^* & \alpha^* & \alpha^*\\
1 & 1 & 1 & 3 & 3 & 3 & 3 & 3 & \alpha^* & \alpha^* & \alpha^* & \alpha & \alpha & \alpha\\
3 & 3 & 3 & 9 & -3 & -3 & -3 & -3 & \cdot & \cdot & \cdot & \cdot & \cdot & \cdot\\
3 & 3 & 3 & -3 & 9 & -3 & -3 & -3 & \cdot & \cdot & \cdot & \cdot & \cdot & \cdot\\
3 & 3 & 3 & -3 & -3 & 9 & -3 & -3 & \cdot & \cdot & \cdot & \cdot & \cdot & \cdot\\
3 & 3 & 3 & -3 & -3 & -3 & -3 & 9 & \cdot & \cdot & \cdot & \cdot & \cdot & \cdot\\
3 & 3 & 3 & -3 & -3 & -3 & 9 & -3 & \cdot & \cdot & \cdot & \cdot & \cdot & \cdot\\
\hline
4 & \alpha & \alpha^* & \cdot & \cdot & \cdot & \cdot & \cdot & 4 & \alpha^* & \alpha & 4 & \alpha^* & \alpha\\
4 & \alpha & \alpha^* & \cdot & \cdot & \cdot & \cdot & \cdot & \alpha^* & \alpha & 4 & \alpha & 4 & \alpha^*\\
4 & \alpha & \alpha^* & \cdot & \cdot & \cdot & \cdot & \cdot & \alpha & 4 & \alpha^* & \alpha^* & \alpha & 4\\
4 & \alpha^* & \alpha & \cdot & \cdot & \cdot & \cdot & \cdot & 4 & \alpha & \alpha^* & 4 & \alpha & \alpha^*\\
4 & \alpha^* & \alpha & \cdot & \cdot & \cdot & \cdot & \cdot & \alpha^* & 4 & \alpha & \alpha & \alpha^* & 4\\
4 & \alpha^* & \alpha & \cdot & \cdot & \cdot & \cdot & \cdot & \alpha & \alpha^* & 4 & \alpha^* & 4 & \alpha
\end{array}
\right),
\end{equation}

\begin{align}
    T=\operatorname{diag}(1,1,1,1,1,1,-1,-1,-1,e^{\frac{2\pi i}{3}}, e^{-\frac{2\pi i}{3}},1,  e^{-\frac{2\pi i}{3}},  e^{\frac{2\pi i}{3}}),
\end{align}
where $\al=2i(i+\sqrt{3})$ and the dots denote zeros.

According to the theorem above, the $(\mathbb{Z}_3,2)$ non-symmetrically braided category, as the subcategory of $\mathfrak{D}(A_4)$, consists of simple objects $\{A,B,C,G\}$ or $\{A,B,C,H\}$. Objects $G$ and $H$ share the same $S$ and $T$ matrix elements, indicating a $G-H$ exchange topological symmetry in $\mathfrak{D}(A_4)$, and they generate the same braided fusion subcategory. 

By the definition of $S$ matrix, $S_{ab}=\frac{1}{D}\operatorname{Tr}(R^{ba}R^{ab})$ where $D$ is the total quantum dimension, the $S$ matrix of the subcategory is the corresponding submatrix, up to a normalization factor:
\begin{align}\label{S ABCG}
S_{\text{sub}} = 
\frac{1}{\sqrt{12}}
\begin{pmatrix}
1 & 1 & 1 & 3 \\
1 & 1 & 1 & 3 \\
1 & 1 & 1 & 3 \\
3 & 3 & 3 & -3
\end{pmatrix}.
\end{align}

\subsection{Identification as Fusion Category}\label{near gp 2}
In this subsection we show that the subcategory $\mathcal{D}$ in Section \ref{sec: su3 wzw} is the non-symmetrically braided $(\mathbb{Z}_3,2)$ described above, and we prove that $\mathcal{D}$ is isomorphic to $\operatorname{Rep}(A_4)$ as a fusion category after forgetting braiding.

The $S$ matrix of $SU(3)_3$ WZW model is
\begin{align}\label{S WZW}
\begin{split}
& S = \frac{1}{6} 
\begin{pmatrix}
1 & 1 & 1 & 3 & 2 & 2 & 2 & 2 & 2 & 2 \\
1 & 1 & 1 & 3 & -1 & -1 & -1 & -1 & -1 & -1 \\
1 & 1 & 1 & 3 & -1 & -1 & -1 & -1 & -1 & -1 \\
3 & 3 & 3 & -3 & 0 & 0 & 0 & 0 & 0 & 0 \\
2 & -1 & -1 & 0 & x & y & z & x & z & y \\
2 & -1 & -1 & 0 & y & z & x & y & x & z \\
2 & -1 & -1 & 0 & z & x & y & z & y & x \\
2 & -1 & -1 & 0 & x & y & z & x & z & y \\
2 & -1 & -1 & 0 & z & x & y & z & y & x \\
2 & -1 & -1 & 0 & y & z & x & y & x & z
\end{pmatrix}\\
&+\frac{i}{6}
\begin{pmatrix}
0 & 0 & 0 & 0 & 0 & 0 & 0 & 0 & 0 & 0 \\
0 & 0 & 0 & 0 & \sqrt{3} & \sqrt{3} & \sqrt{3} & -\sqrt{3} & -\sqrt{3} & -\sqrt{3} \\
0 & 0 & 0 & 0 & -\sqrt{3} & -\sqrt{3} & -\sqrt{3} & \sqrt{3} & \sqrt{3} & \sqrt{3} \\
0 & 0 & 0 & 0 & 0 & 0 & 0 & 0 & 0 & 0 \\
0 & \sqrt{3} & -\sqrt{3} & 0 & a & b & c & a & c & b \\
0 & \sqrt{3} & -\sqrt{3} & 0 & b & c & a & b & a & c \\
0 & \sqrt{3} & -\sqrt{3} & 0 & c & a & b & c & b & a \\
0 & -\sqrt{3} & \sqrt{3} & 0 & a & b & c & a & c & b \\
0 & -\sqrt{3} & \sqrt{3} & 0 & c & a & b & c & b & a \\
0 & -\sqrt{3} & \sqrt{3} & 0 & b & c & a & b & a & c
\end{pmatrix},
\end{split}
\end{align}
where
\begin{align}
\begin{split}
x &= 2(\cos(\pi/18)+\sin(2\pi/9))/\sqrt{3},\\
y &= -2(\cos(\pi/18)+\sin(\pi/9))/\sqrt{3},\\
z &= 2(\sin(\pi/9)-\sin(2\pi/9))/\sqrt{3},\\
a &= -2(\cos(2\pi/9)+\sin(\pi/18))\sqrt{3},\\
b &= -2(\cos(\pi/9)-\sin(\pi/18))/\sqrt{3},\\
c &= -2(\cos(\pi/9)+\cos(2\pi/9))\sqrt{3}.
\end{split}
\end{align}
The T matrix is $T_{ij}=\delta_{ij}\exp(2\pi i(h_i-\frac{c}{24}))$\footnote{Here $c=4$ is the central charge of $SU(3)_3$ WZW CFT.}, where the conformal weights $h_i$'s are given in \eqref{verlinde in su(3)3}.

The subcategory $\mathcal{D}$ consisting of the first 4 objects ($1,\tau,\tau^2$ and $X$) inherits the braiding from the modular tensor category of total 10 Verlinde lines, with the matrix $S$ given by the upper left block of \eqref{S WZW}. Note that $S_{XX}=-\frac{1}{2}$ is negative, meaning that $\mathcal{D}$ cannot be \emph{symmetrically} braided. Furthermore, the fusion rules given in Section \ref{sec: su3 wzw} are of $(\mathbb{Z}_3,2)$ type. Hence, the category $\mathcal{D}$ satisfied the condition in the theorem in the above subsection, and is exactly the braided category described above
\begin{align}
\begin{split}
    &\mathcal{D} \xhookrightarrow{\text{inclusion}}\mathfrak{D}(A_4)\\
    &\operatorname{Irr}(\mathcal{D})= \{ A, B, C, G\} \quad \text{or} \quad \{ A, B, C, H\}.
\end{split}
\end{align}
As a sanity check, the $S$ matrix of $\mathcal{D}$, the the upper left block of \eqref{S WZW} up to a normalization factor, is the same as \eqref{S ABCG}.

With this inclusion, we can show that $\mathcal{D}$ is isomorphic to $\operatorname{Rep}(A_4)$ as a fusion category after forgetting braiding.

The modular tensor category $\mathfrak{D}(A_4)$ accepts a tensor functor to $\operatorname{Rep}(A_4)$, where it acts on objects as the induced representation:
\begin{align}
\begin{split}
    \mathfrak{D}(A_4)&\xrightarrow[]{F} \operatorname{Rep}(A_4)\\
    ([g],\pi)&\mapsto\rho.
\end{split}
\end{align}
Here $\rho$ is a (not necessarily irreducible) representation of $A_4$ induced from the representation $\pi$ of $Z_{g}$, where $Z_{g}$ is treated as the subgroup of $A_4$. Physically, we can think of this functor as the bulk ($A_3$ gauge theory)-to-flux-boundary functor.

By restricting to $\{A,B,C,G\}$ (or $\{A,B,C,H\}$), the functor is an isomorphism on objects
\begin{align}
    \{A,B,C,G\}\mapsto\{1,1',1'',3\},
\end{align}
where we label representations of $A_3$ by their dimensions. Since $F$ is a tensor functor, we arrive at the conclusion that after forgetting the braiding structure on $\mathcal{D}$, $\mathcal{D}$ is isomorphic to $\operatorname{Rep}(A_4)$ as a fusion category.

\bibliographystyle{JHEP}
\bibliography{ref}

\end{document}